\documentclass[twocolumn,preprintnumbers,superscriptaddress,nofootinbib,aps,prd,floatfix]{revtex4}

\usepackage{footmisc,multirow}
\usepackage{subfigure,enumerate}
\usepackage{amsmath,amssymb}

\usepackage{graphicx}

\usepackage{color}

\newcommand{\be}{\begin{eqnarray*}}
\newcommand{\ee}{\end{eqnarray*}}
\newcommand{\gl}[1]{(\ref{#1})}

\newcommand{\bee}{\begin{eqnarray}}
\newcommand{\eee}{\end{eqnarray}}
\newcommand{\beeq}{\begin{equation}}
\newcommand{\eeeq}{\end{equation}}
\newcommand{\cp}{${\cal{CP}}$}
\newcommand{\gev}{{\text{GeV}}}

\renewcommand{\vec}{\bf}

\preprint{IPPP/13/20}
\preprint{DCPT/13/40}
\preprint{MPP-2013-97}

\begin{document}

\title{The shape of spins}

\begin{abstract}
  After the discovery of a Higgs-like particle at the LHC, the
  determination of its spin quantum numbers across different channels
  will be the next step in arriving at a more precise understanding of
  the new state and its role in electroweak symmetry breaking. Event
  shape observables have been shown to provide extremely sensitive
  observables for the discrimination of the scalar Higgs boson's
  \cp~quantum numbers as a consequence of the different radiation
  patterns of Higgs production via gluon fusion vs. weak boson fusion
  in the $pp\to X+2j$ selection. We show that a similar strategy
  serves to constrain the spin quantum numbers of the discovered
  particle as a function of the involved couplings. We also discuss
  the prospects of applying a similar strategy to future discoveries
  of Higgs-like particles.
\end{abstract}

\author{Christoph Englert} \email{christoph.englert@durham.ac.uk}
\affiliation{Institute for Particle Physics Phenomenology, Department
  of Physics,\\Durham University, DH1 3LE, United Kingdom}

\author{Dorival Gon\c{c}alves} \email{dorival@mpp.mpg.de}
\affiliation{Institut f\"ur Theoretische Physik, Universit\"at
  Heidelberg, 69120 Heidelberg, Germany}
\affiliation{Max-Planck-Institut f\"ur Physik, F\"ohringer Ring 6,
  80805 M\"unchen, Germany}

\author{Graeme Nail} \email{graeme.nail@durham.ac.uk}
\affiliation{Institute for Particle Physics Phenomenology, Department
  of Physics,\\Durham University, DH1 3LE, United Kingdom}

\author{Michael Spannowsky} \email{michael.spannowsky@durham.ac.uk}
\affiliation{Institute for Particle Physics Phenomenology, Department
  of Physics,\\Durham University, DH1 3LE, United Kingdom}

\maketitle


\section{Introduction}
After the discovery of a Standard Model Higgs boson-like
particle~\cite{orig} at the LHC~\cite{:2012gk,:2012gu}, the
measurement its spin is the next step in arriving at a more 
complete picture of this discovery. There is a 
theoretical prejudice from Lorentz invariance against spin
$J=1$~\cite{landauyang} as the particle is observed in the decay to
photons, which leaves scalar $J=0$ as the well-defined option in terms
of our current understanding of perturbative Quantum Field Theory.

There is a known caveat in analyzing spin hypotheses $J\geq 2$ that
arises when we investigate tensor particles and beyond. As a matter of
fact, there is no well-behaved QFT which predicts the interactions of
such a state with SM matter from first principles. In particular,
there are certain indirect constraints on the spin $J=2$ options if we
take into account the non-observation of large excesses in $VV+2j$
final states ($V=W^\pm,Z$) at the LHC so far, while there is
consistency in $X\to VV$ with the SM within errors. The latter implies
that the observed particle is involved in the unitarization of
$V_LV_L$ scattering and probably provides the dominant share to the
saturation of the unitarity sum rules. In simple realizations, this
cannot be achieved with a spin 2 particle~\cite{juergen} and the
worsened unitarity problem in longitudinal gauge boson scattering
would manifest in a large cross section in the $VV+2j$ final state at
large invariant masses.

On the other hand, we can perform spin analyses beyond indirect
constraints in model-independent ways in the fully reconstructible
final states $X\to
ZZ,\gamma\gamma$~\hbox{\cite{lykken,Choi:2012yg,Ellis:2012wg,cp}}.
Many of the direct measurement analysis strategies originate from
similar questions addressed in hadron
physics~\cite{Cabibbo:1965zz}. Doing so, one typically treats the
$X$~decay independent from~$X$ production.\footnote{The simulation of
  such final states, however, needs to include the full matrix element
  because, {\it e.g.}, for a graviton-like object the only source of
  deviation is the propagator, see also~\cite{Banerjee:2012ez}.}
Indeed, recent LHC measurements along these lines seem to favor
$J^{\cal{CP}}=0^+$ searches~\cite{exispin,djou}.

However, treating the resonance's decay independent from its
production does not allow one to draw a more complete picture of Higgs
couplings because momentum dependencies are typically encoded in
off-shell effects that cannot be studied in this way. It is precisely
the momentum dependence of higher dimensional operators that leaves
footprints in the $X+2j$ channel~\cite{vera}, {\it i.e.}, the $t$
channel gauge bosons in the weak boson fusion (WBF) topologies are 
always virtual. In this sense, adapted search strategies for the $X+2j$ 
selection do not only provide additional sensitivity, which 
can be used in a global spin hypothesis test across various channels, 
but also include orthogonal information that cannot be accessed 
via more traditional spin measurements.

In this letter we show that the global energy flow structure that
follows from typical representatives of alternative spin structures
provides a highly sensitive observable to study these properties. We
select combinations of couplings, right from the beginning, that lead
to a SM Higgs-like phenomenology. As
Refs.~\cite{Frank:2012wh,Englert:2012xt,Djouadi:2013yb} explain, the
``tagging'' jet kinematics in $X+2j$ final states can be a strong
discriminant for the spin of the produced particle~$X$. It should be
noted that this typically results from the involved
(higher-dimensional) operator structures, which are determined by the
spin hypotheses. With this in mind, we specifically analyze spin 2
models that have $p_T$ distributions similar to the SM
Higgs~\cite{Englert:2012xt}. In doing so, we complement the analyzes
of~\cite{Frank:2012wh,Englert:2012xt,Djouadi:2013yb} by answering how
much sensitivity hides beyond the tagging jet level and how it carries
over to experimental reality.

We will also investigate the strategy's prospects for heavier
``Higgs'' masses. This latter point is motivated by the fact that
similar questions, as to those we currently face for the 125 GeV particle,
will arise if additional Higgs-like states are discovered in the
future. Such states are predicted by many extensions of the SM Higgs
sector.

\begin{figure}[!t]
  \centering
  \includegraphics[width=0.45\textwidth]{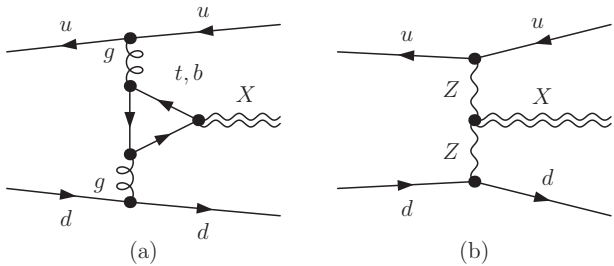}
  \caption{\label{fig:feyngraphs} Sample Feynman diagrams that
    contribute to $X+2j$ production via gluon fusion (a) and weak boson
    fusion (b). We do not show the $X$ decay.}
\end{figure}

\subsubsection*{Event shapes as electroweak-sensitive observables}
The azimuthal angle between the two tagging jets in the $pp\to X+2j$
selection $\Delta\Phi_{jj}$~\cite{Andersen:2008,ken,kinem,paco,kinem2}
defined according to rapidity $y$
\begin{equation}
  \label{eq:phijj}
  \Delta\Phi_{jj} = \phi(p_>) - \phi(p_<)\,,
\end{equation}
where $p^\mu_{\lessgtr} =\sum_{{{j\in \{\rm{jets}}} :\; y_j{\lessgtr
  }y_X\}} p^\mu_j$, is known to be a highly sensitive observable to
the \cp~quantum numbers of the produced $X$~state. This finding is not
limited to the WBF channels~\cite{Plehn:2001nj}, but is known to also
work in the gluon fusion channel~\cite{Klamke:2007cu,kinem2}. The
latter production mechanism can give rise to \cp~odd Higgs production
via tree-level \cp~odd couplings to the heavy fermion sector,
Fig.~\ref{fig:feyngraphs}. Such a state is typically present in any
non-singlet Higgs sector extension that feature fields transforming in
non-trivial representations under SU(2)$_L$. In the light of recent
measurements, the fields of these extensions need to be heavier, with
suppressed cross sections.

Another way to understand the sensitivity encoded in $\Delta\Phi_{jj}$
is that the amplitude as a whole is sensitive to the \cp~quantum
numbers. Hence, any additional \cp-preserving QCD leg that is attached
to diagrams in Fig.~\ref{fig:feyngraphs} will still give rise to an
amplitude which encodes the \cp-specific properties reflected in
$\Delta\Phi_{jj}$ for two-jet configurations. As a result, the entire
QCD activity that results from the hard interactions in
Fig.~\ref{fig:feyngraphs} can be considered a probe of the produced
state~$X$.\footnote{In principle this argument extends also to the
  soft coherent radiation down to the hadronization scale. These
  effects are however subleading.} Finding the ``proper'' jets of
Fig.~\ref{fig:feyngraphs} that reflect the nature of the produced
state in a multi-jet environment amounts to a combinatorial and
quantum-interference--governed problem; this results in reduced
sensitivity in the $\geq 3j$ selection~\cite{Andersen:2008}.

A way to circumvent this was outlined in Ref.~\cite{Englert:2012xt}:
Since QCD radiation implies energy-momentum flow, the entire energy
distribution in the detector (upon reconstructing and removing~$X$
from the list of calorimeter hits) can be expected to provide a
superior discriminant compared to $\Delta\Phi_{jj}$ in an inclusive
selection. The energy momentum flow of an LHC event is commonly
quantified by means of hadronic event shape
observables~\cite{evtshapes}.\footnote{See {\it e.g.}
  Ref.~\cite{rivet} for publicly available implementations within the
  {\sc{Rivet}} analysis package.} Indeed, Ref.~\cite{Englert:2012ct}
found an increase in sensitivity that follows from investigating event
shapes for discrete \cp~measurements.\footnote{Since the sensitivity
  does not follow from a specific angular distribution
  $\Delta\Phi_{jj}$ still remains the observable of choice for mixed
  \cp~states, which can be straightforwardly extracted by fitting
  trigonometrical functions for an essentially background-free
  selection~\cite{Plehn:2001nj,Klamke:2007cu}. This procedure only
  becomes available at high integrated luminosities.}  The interplay
between event shapes and Higgs physics was further studied in
Ref.~\cite{Bernaciak:2012nh}.

Recently in Refs.~\cite{Djouadi:2013yb,Englert:2012xt,Frank:2012wh} a
substantial discriminative power was revealed in the $pp\to X+2j$
final state for different spin hypotheses $J(X)$. This sensitivity is
driven by the energy-dependence of operators which mimic the Higgs
boson's interactions. The differences in the observed phenomenology
can be manifold and depends on the specific higher spin scenario that
one investigates. However, a rather generic finding is that spin 1 and
2 operators tend to populate the central region of the detector, thus
leading to a departure from a WBF-like signature; consequentially
central jet vetos~\cite{early,vetos} need to be relaxed to be
sensitive to such an event topology. This means that backgrounds need
to be suppressed by a combination of stiff $b$ vetos~\cite{atltag} and
state-of-the-art signal vs. background ($S/B$) discriminators, such as
the matrix element method~\cite{Freitas:2012uk}, depending on the
final state.\footnote{Another finding
  of~\cite{Djouadi:2013yb,Englert:2012xt} is that the sensitivity
  observed, in the combination of transverse momentum and rapidity
  difference, points to the invariant dijet mass as single
  discriminant.}

In the following we will consider $pp\to X+2j$ with~$X$ decay to fully
leptonic taus for a toy-level signal vs. background study to compare
the performance of various event shape-based observables. The details
of the Higgs reconstruction are inconsequential in this comparison, as
all observables are affected in the same way, and the Higgs candidate
does not enter our analysis apart from reconstructing the signal
within a window cut around the candidate mass of $m_X\simeq
125~\gev$. Hence, we do not include any tau reconstruction
efficiencies that can also vary across the different exclusive tau
decay modes \cite{Chatrchyan:2012vp,exitau}.  We also note that our
analysis strategy is insensitive to the specifics of the ``Higgs''
decay channel, and our methods straightforwardly generalize to other
decay channels such as, {\it e.g.}, the $\gamma\gamma + 2j$ selection.

\section{Analysis Setup}
\label{sec:analysis}
For the purpose of comparability, we closely follow
Ref.~\cite{Englert:2012ct}. We model our signal hypotheses with a
combination of {\sc{MadGraph}}~\cite{madgraph} and
{\sc{Herwig++}}~\cite{herwigpp}. For the simulation of the backgrounds
we generate matched events with {\sc{Sherpa}}~\cite{sherpa} and in the
following limit ourselves to the $t\bar t+$jets and $Z+$jets
backgrounds~\cite{Plehn:2001nj}; normalizing these event samples to
the NNLO~\cite{ttjets,Moch:2008ai} and NLO cross
sections~\cite{vbfnlo,mcfm,zjets}, respectively.

\begin{figure*}[!p]
  \includegraphics[width=0.32\textwidth]{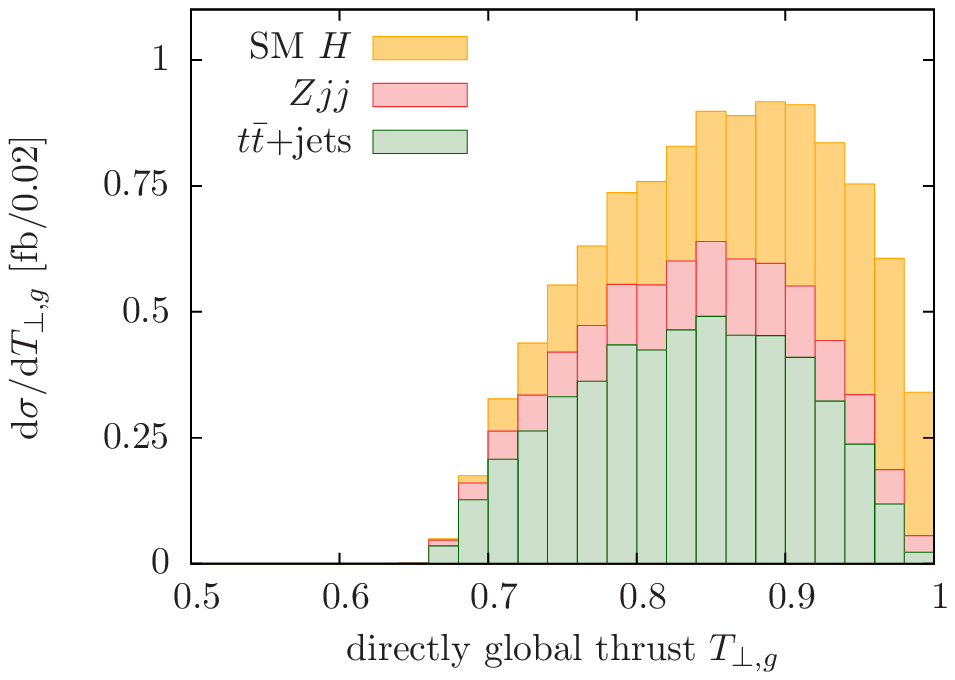}\hfill
  \includegraphics[width=0.32\textwidth]{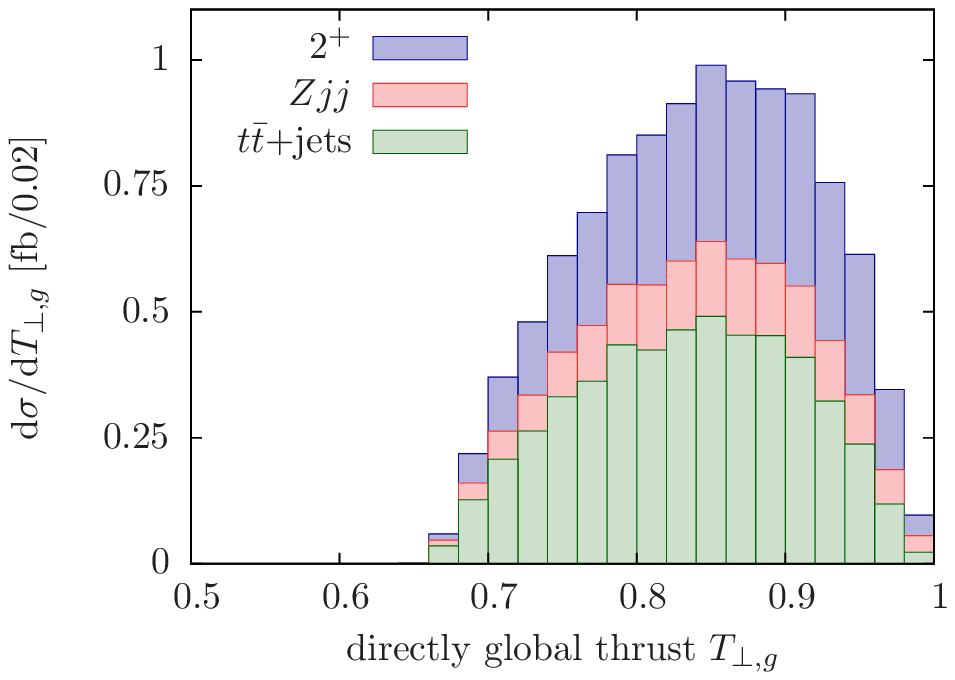}\hfill
  \includegraphics[width=0.32\textwidth]{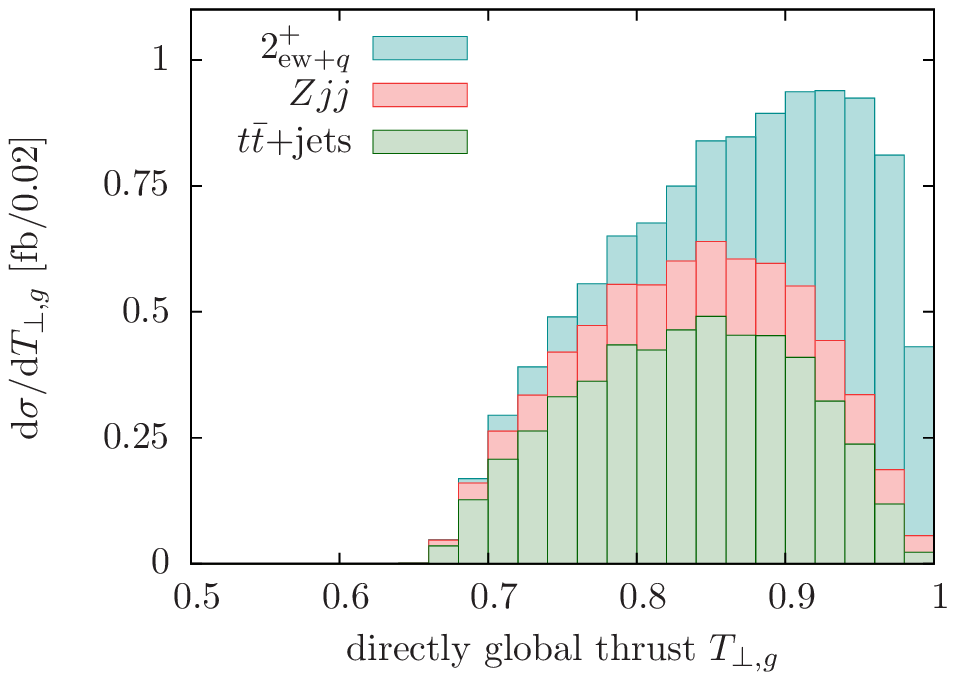}\\[0.6cm]
  \includegraphics[width=0.32\textwidth]{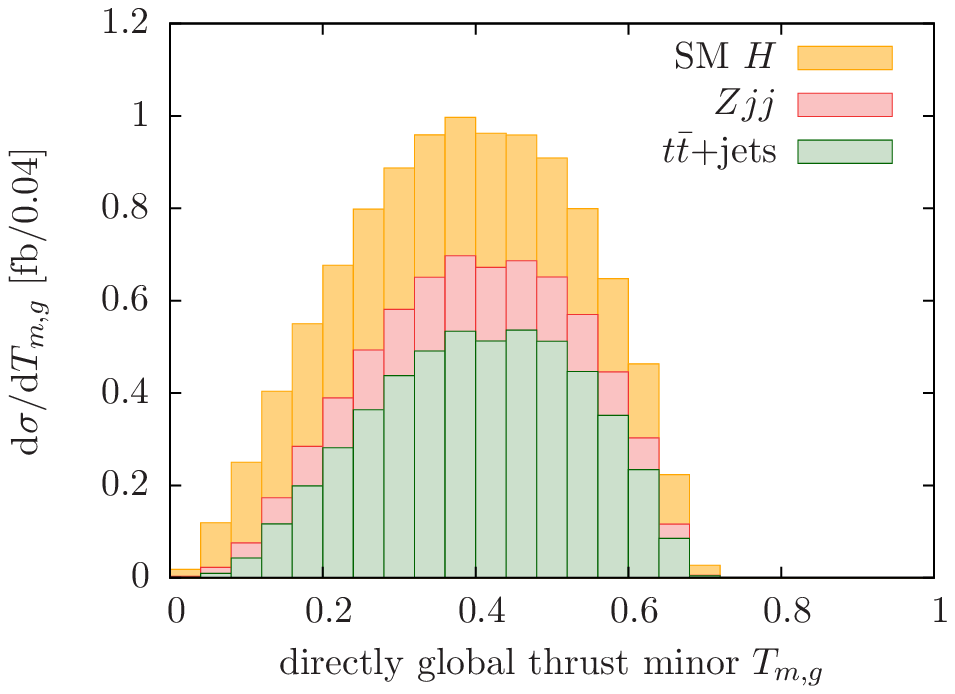}\hfill
  \includegraphics[width=0.32\textwidth]{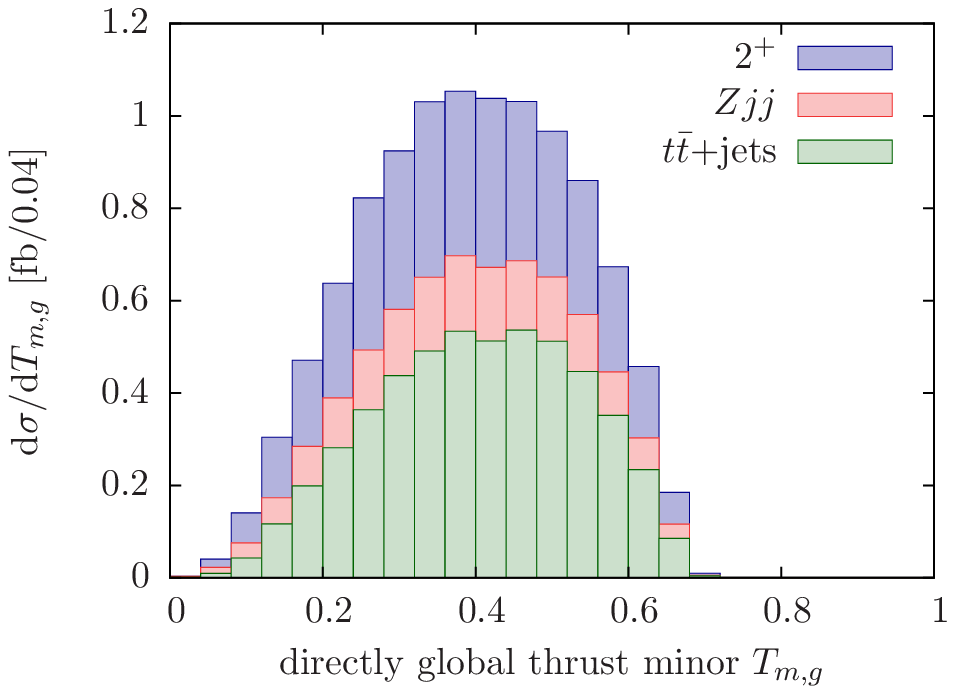}\hfill
  \includegraphics[width=0.32\textwidth]{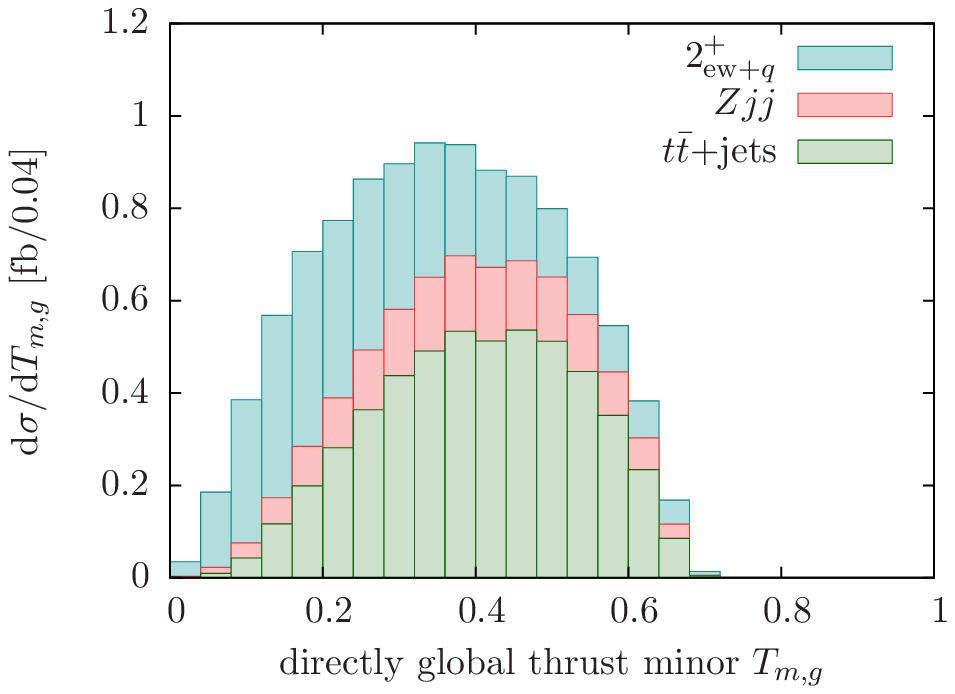}\\[0.6cm]
 \includegraphics[width=0.32\textwidth]{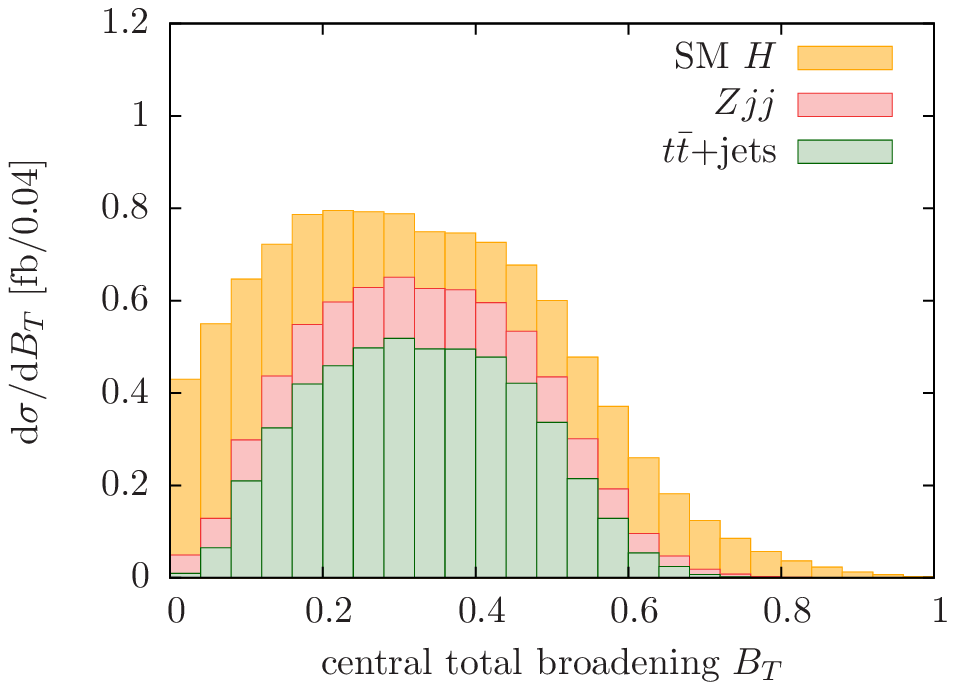}\hfill
 \includegraphics[width=0.32\textwidth]{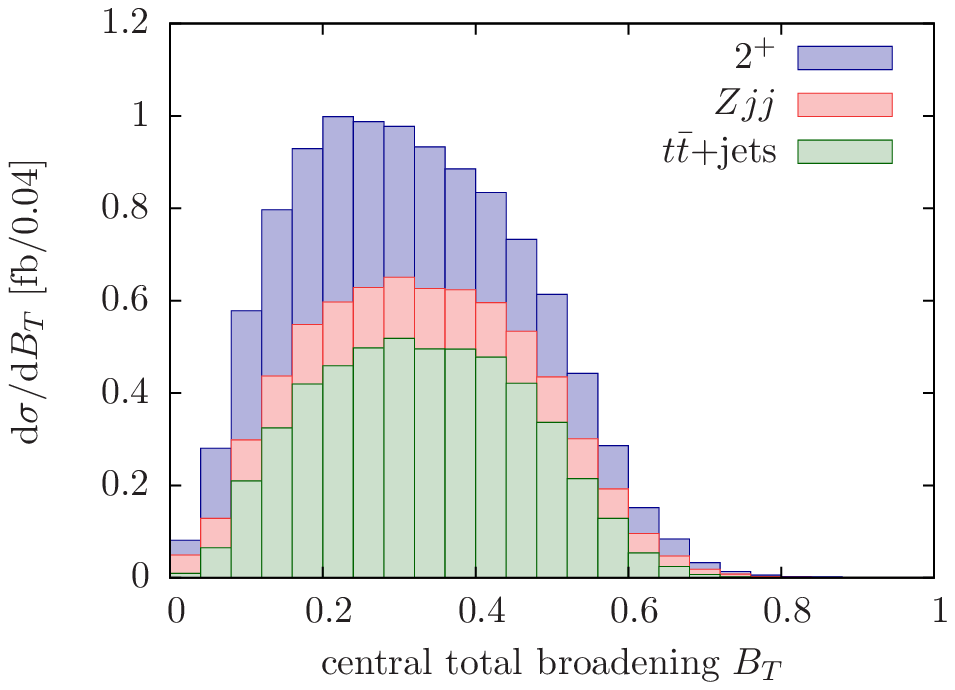}\hfill
 \includegraphics[width=0.32\textwidth]{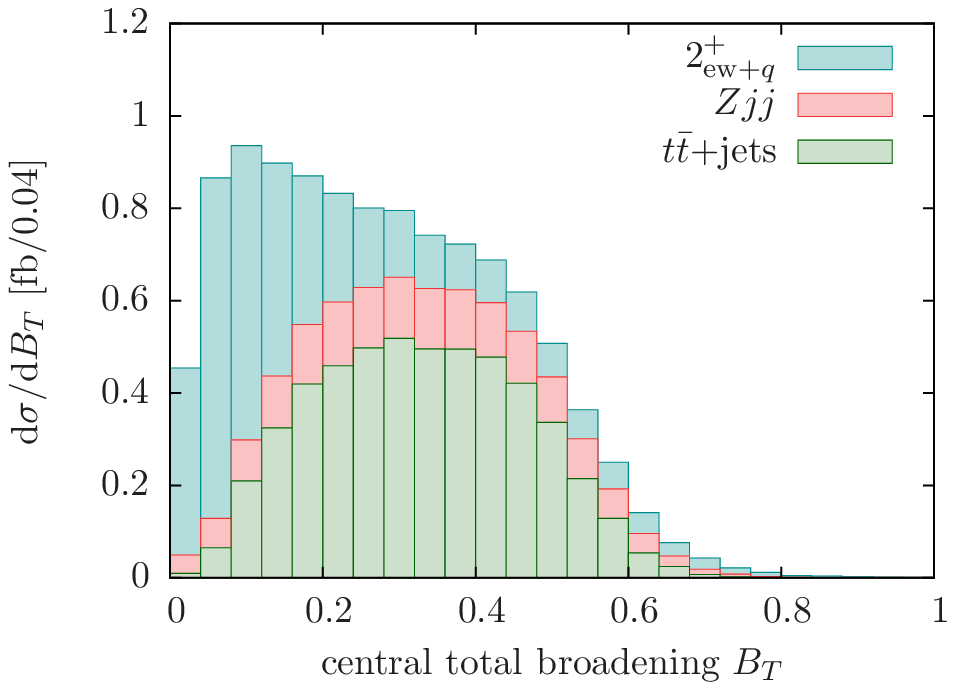}\\[0.6cm]
 \includegraphics[width=0.32\textwidth]{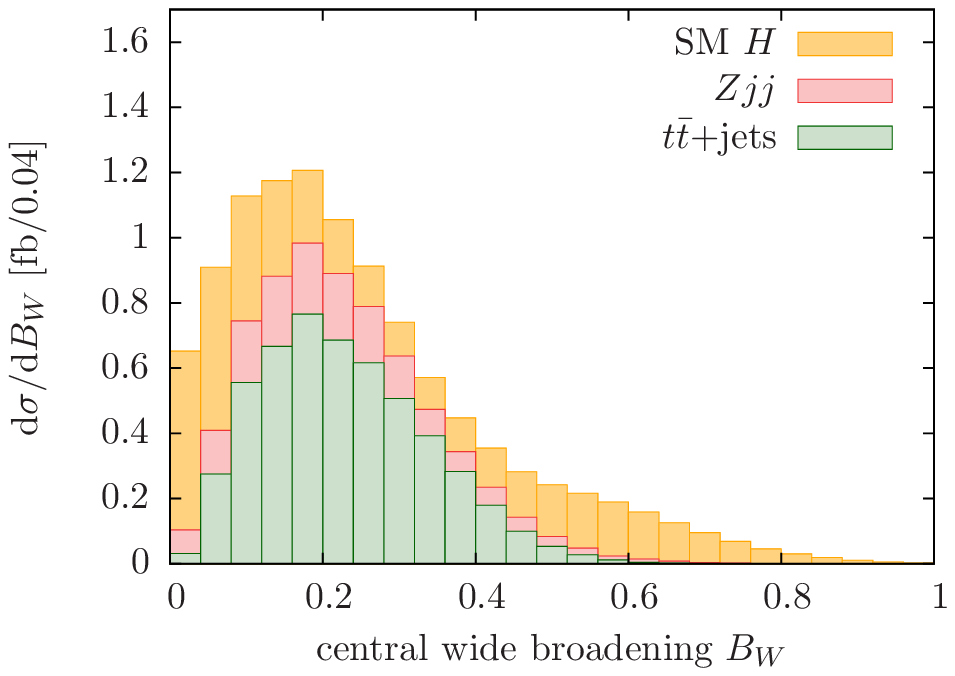}\hfill
 \includegraphics[width=0.32\textwidth]{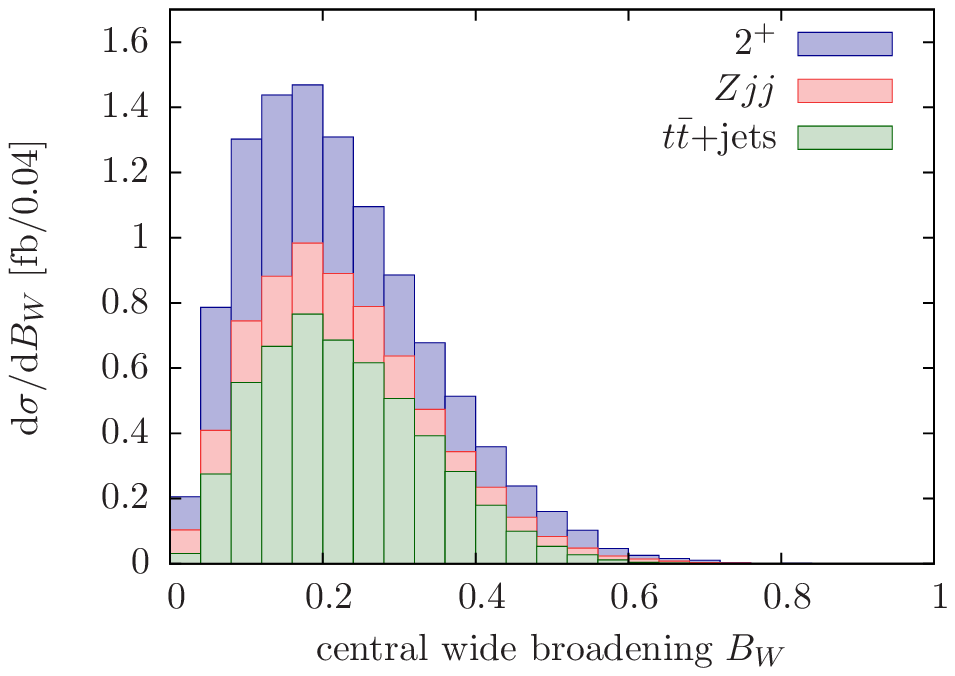}\hfill
 \includegraphics[width=0.32\textwidth]{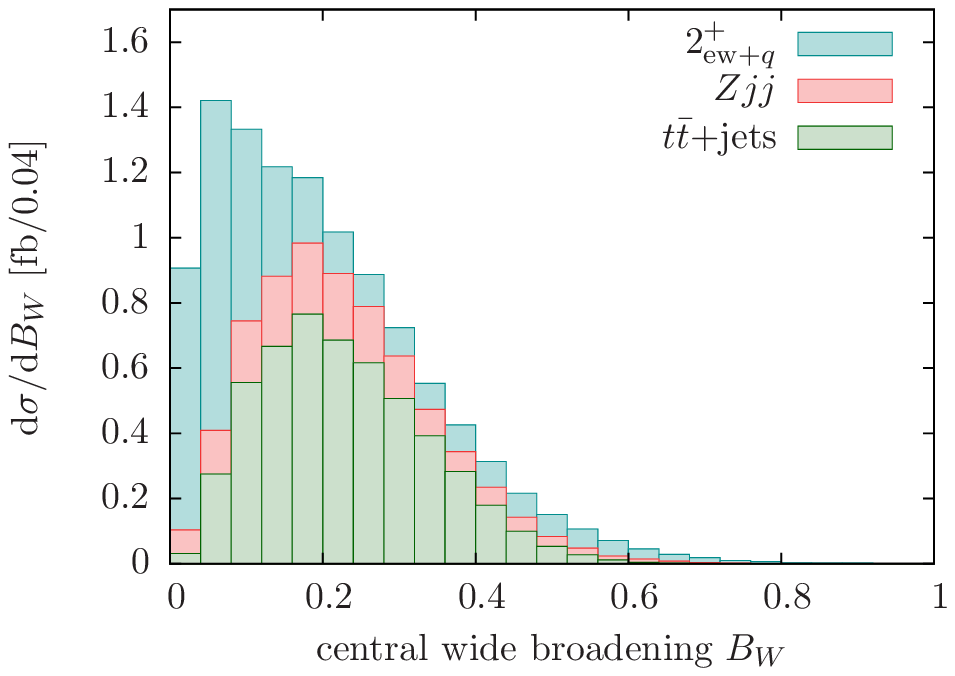}\\[0.6cm]
 \includegraphics[width=0.32\textwidth]{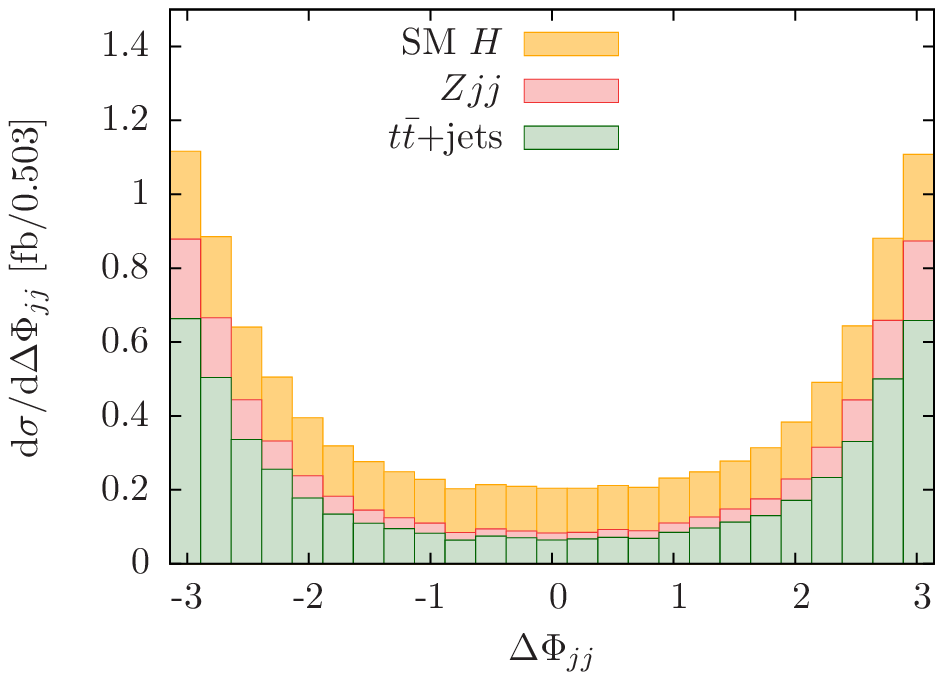}\hfill
  \includegraphics[width=0.32\textwidth]{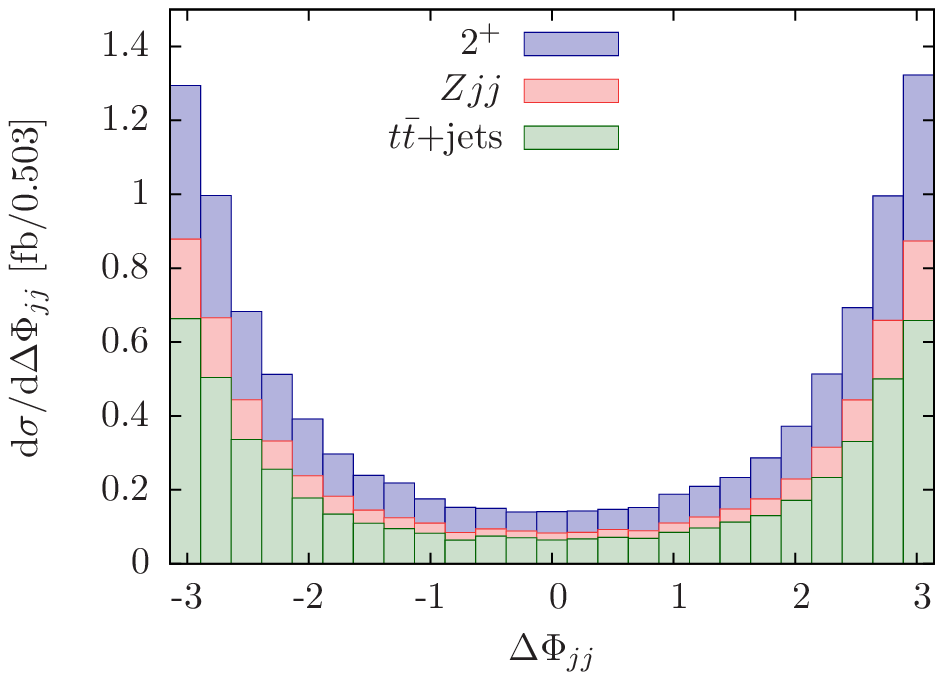}\hfill
  \includegraphics[width=0.32\textwidth]{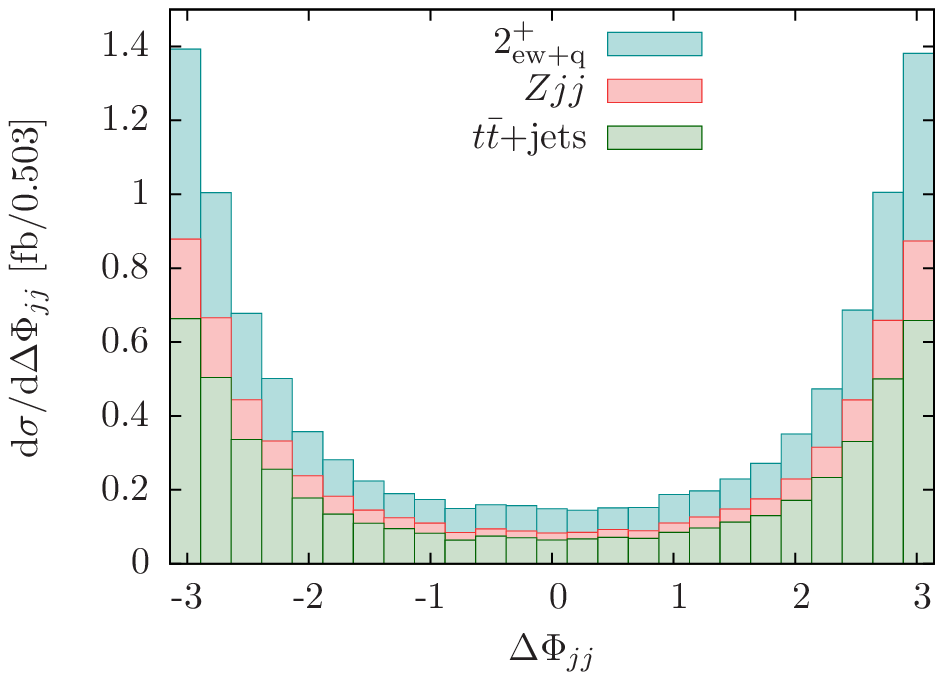}\\[0.2cm]
  \caption{\label{fig:evshapes} Event shape distribution for the
    different event shapes calculated from all particle tracks in
    $|\eta|<4.5$ with $p_{T}\geq 1~\gev$ for the selection~(i). We
    also show $\Delta\Phi_{jj}$.}
\end{figure*}
%
%

\begin{figure*}[!p]
  \includegraphics[width=0.32\textwidth]{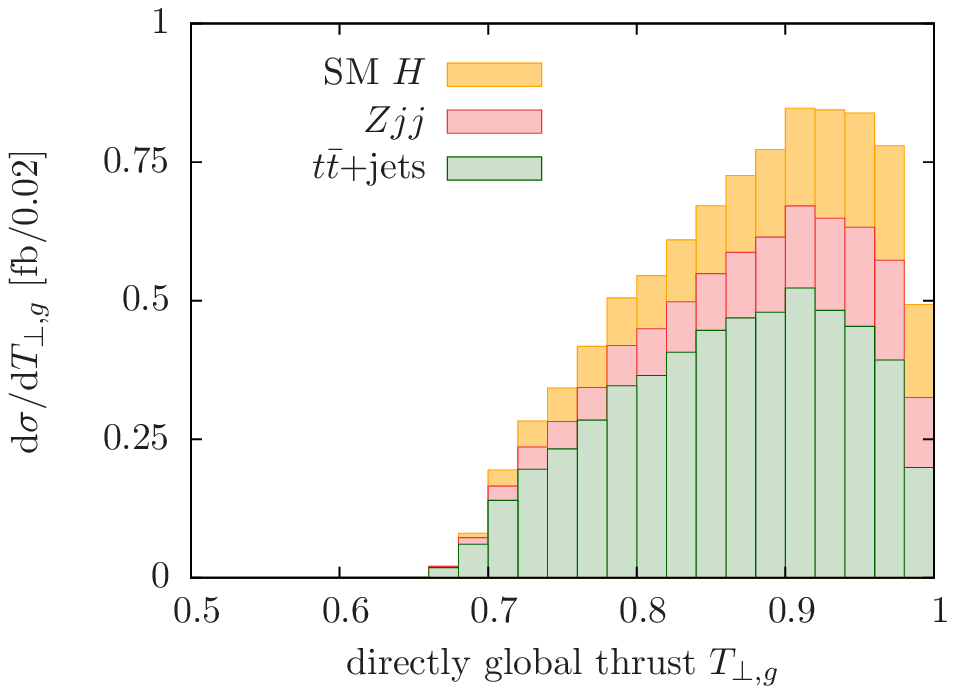}\hfill
  \includegraphics[width=0.32\textwidth]{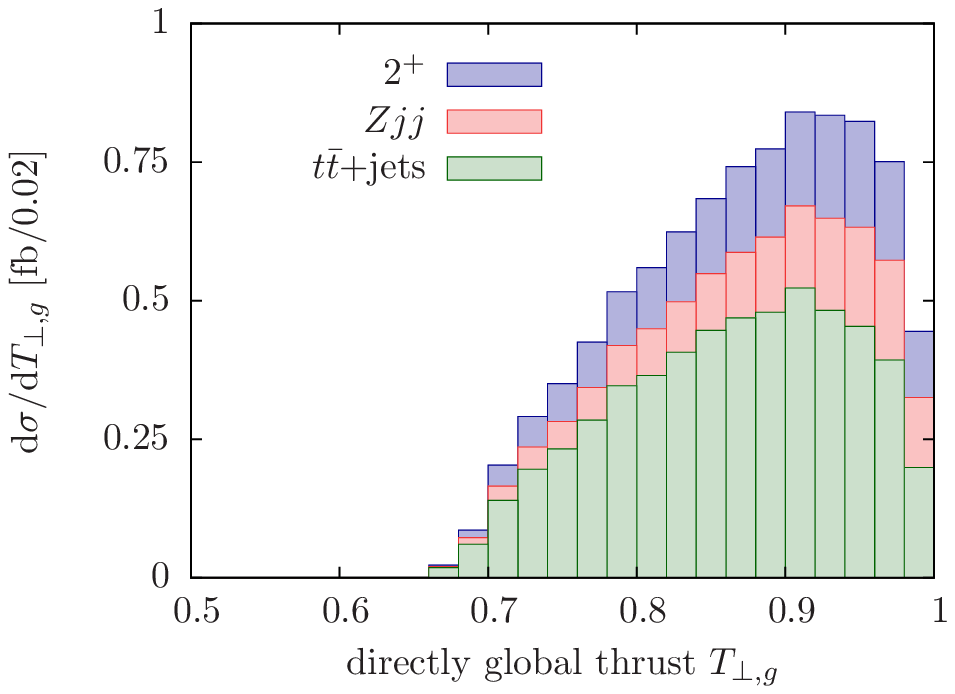}\hfill
  \includegraphics[width=0.32\textwidth]{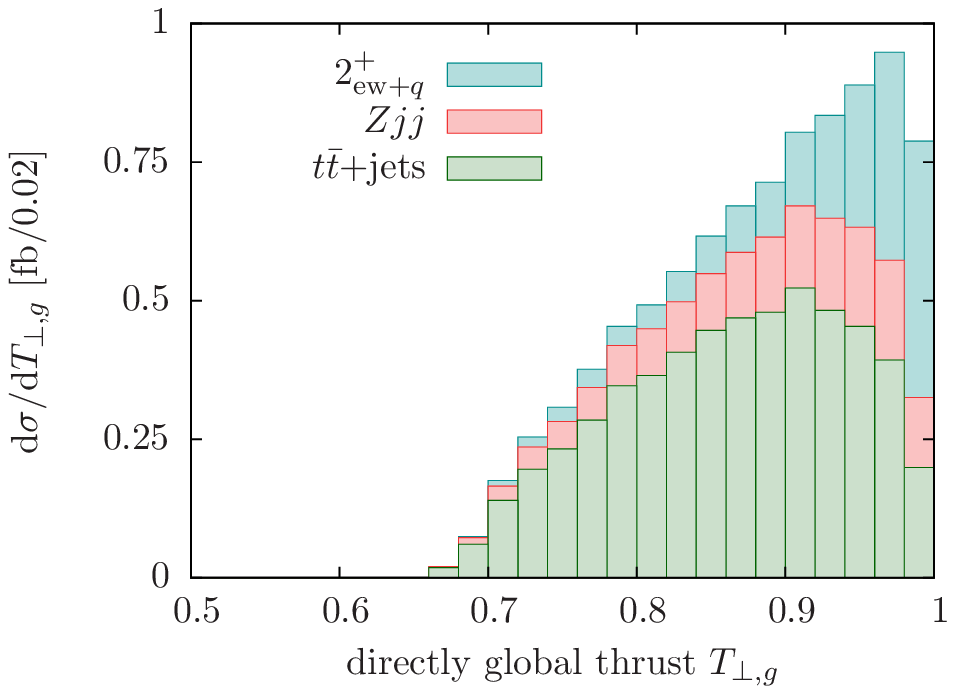}\\[0.6cm]
  \includegraphics[width=0.32\textwidth]{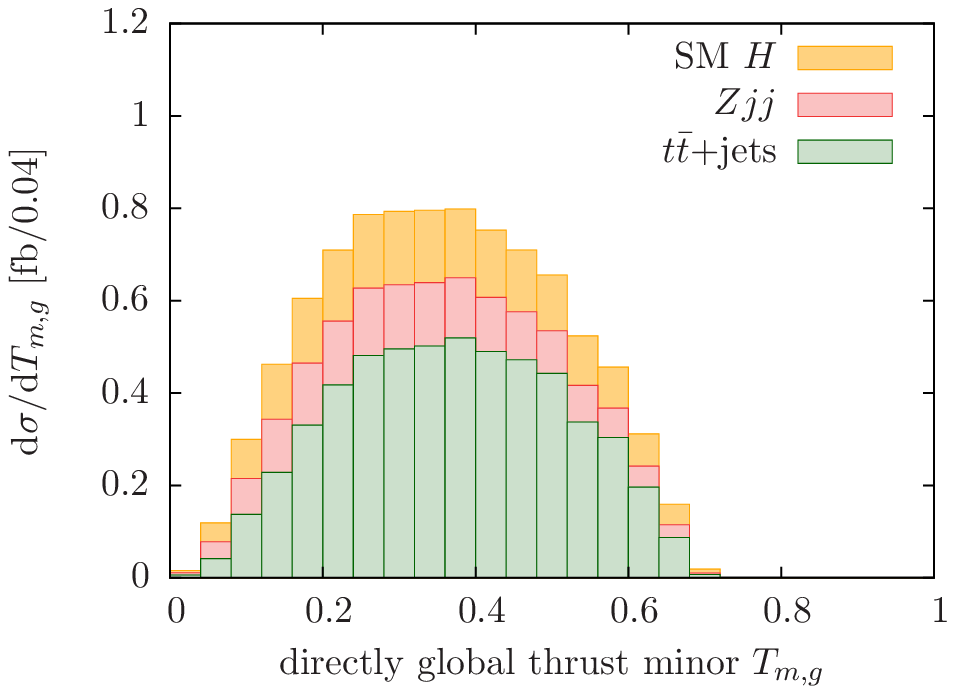}\hfill
  \includegraphics[width=0.32\textwidth]{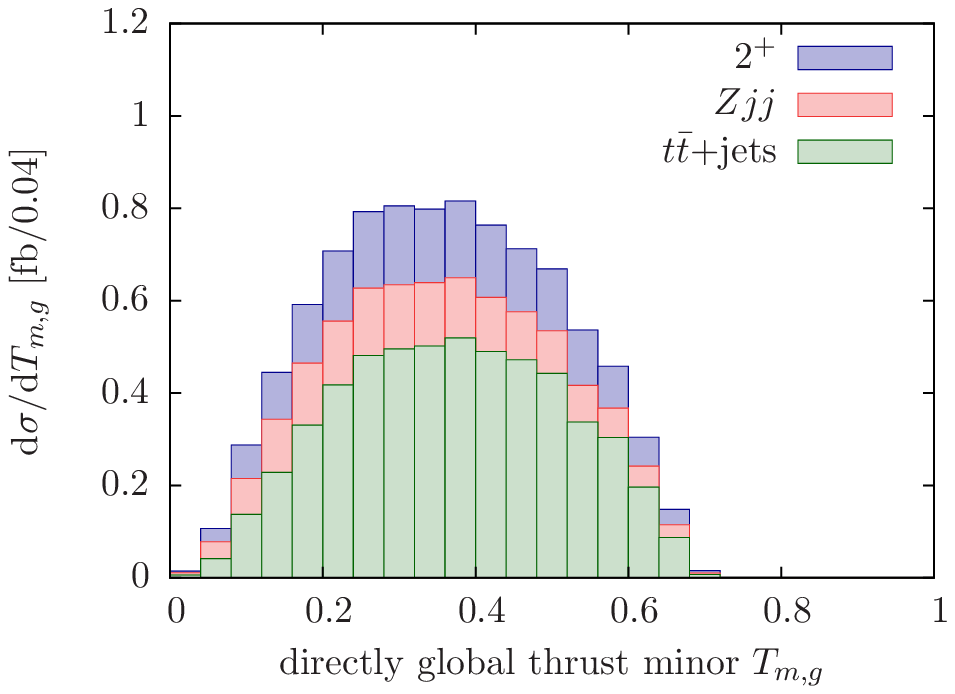}\hfill
  \includegraphics[width=0.32\textwidth]{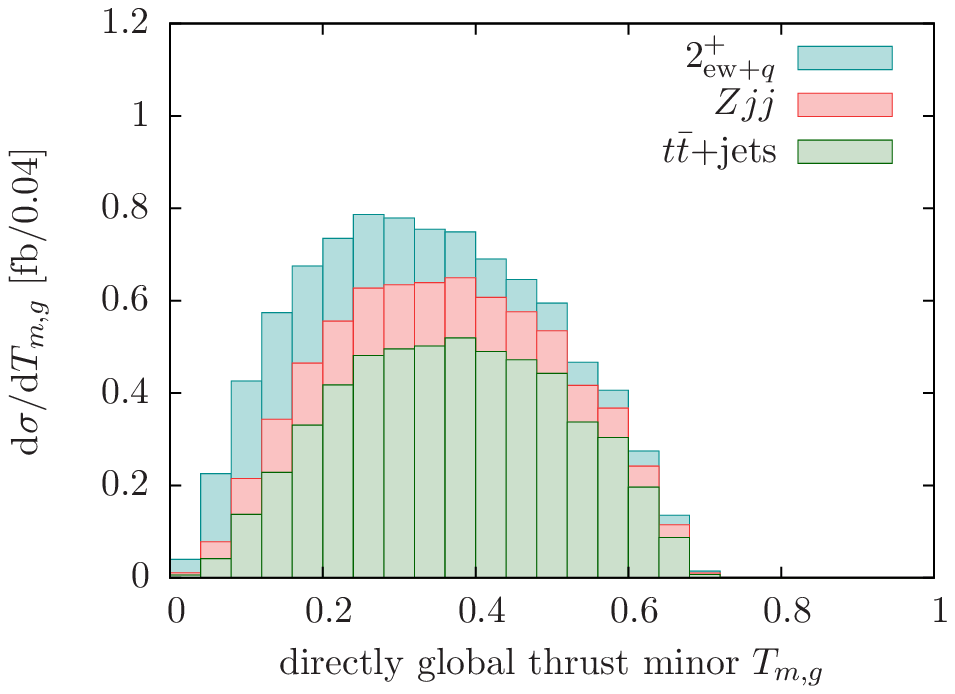}\\[0.6cm]
 \includegraphics[width=0.32\textwidth]{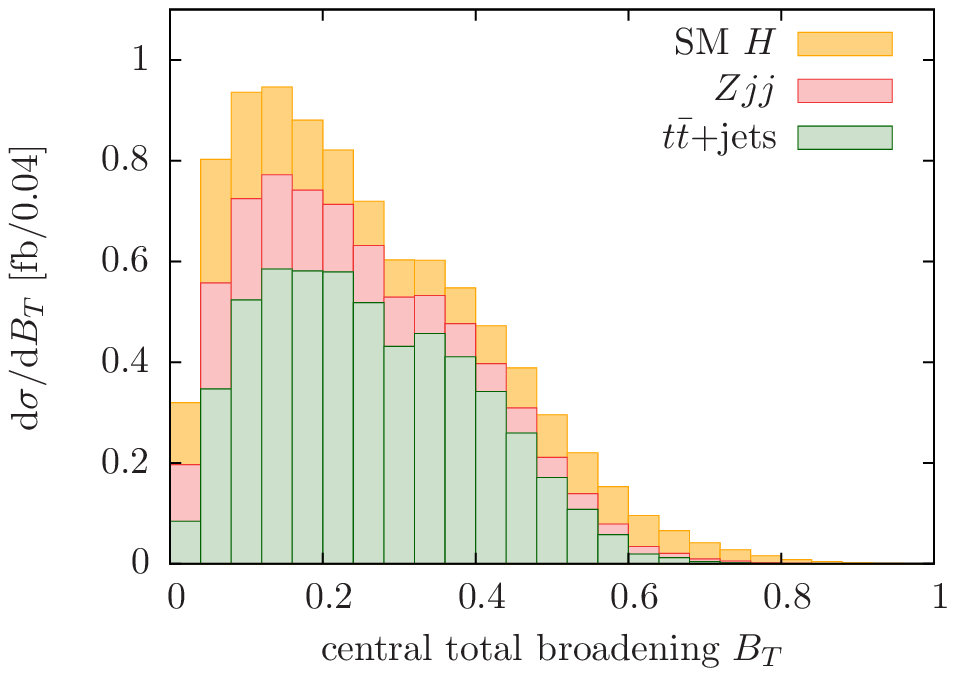}\hfill
 \includegraphics[width=0.32\textwidth]{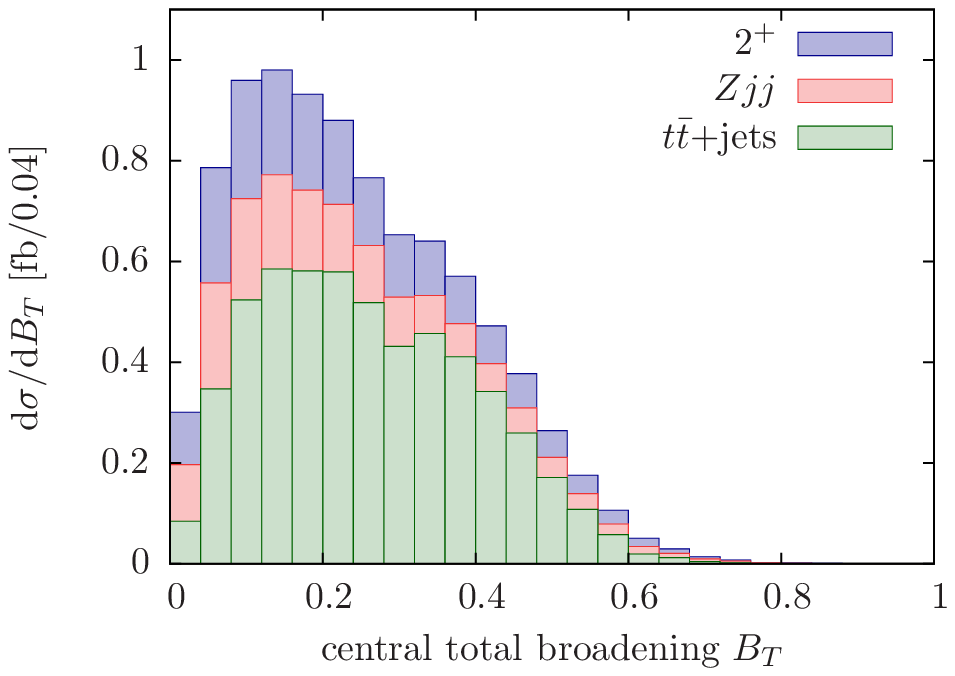}\hfill
 \includegraphics[width=0.32\textwidth]{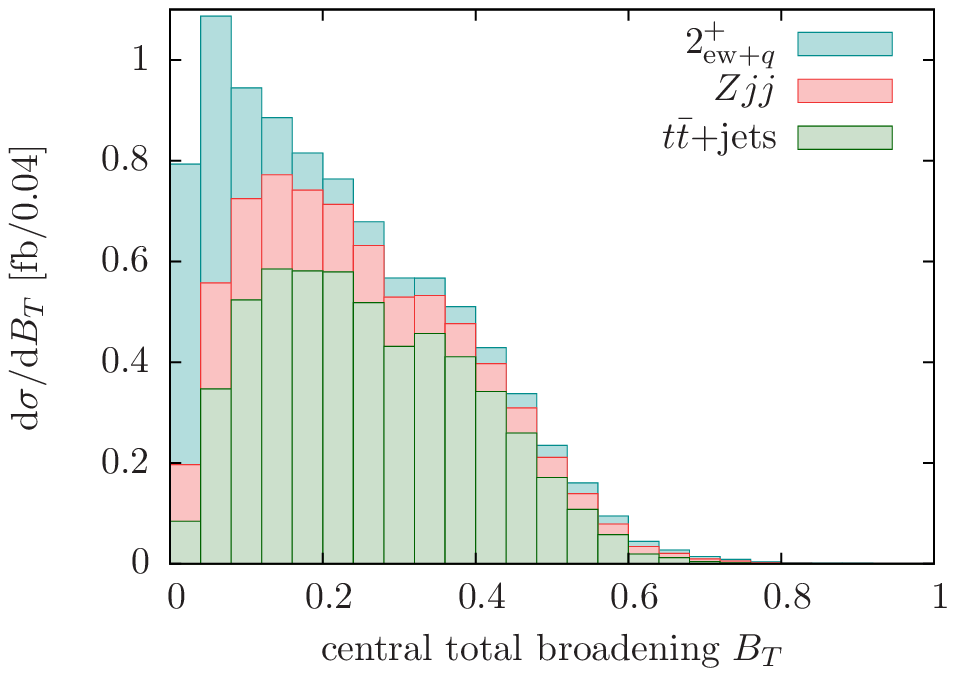}\\[0.6cm]
 \includegraphics[width=0.32\textwidth]{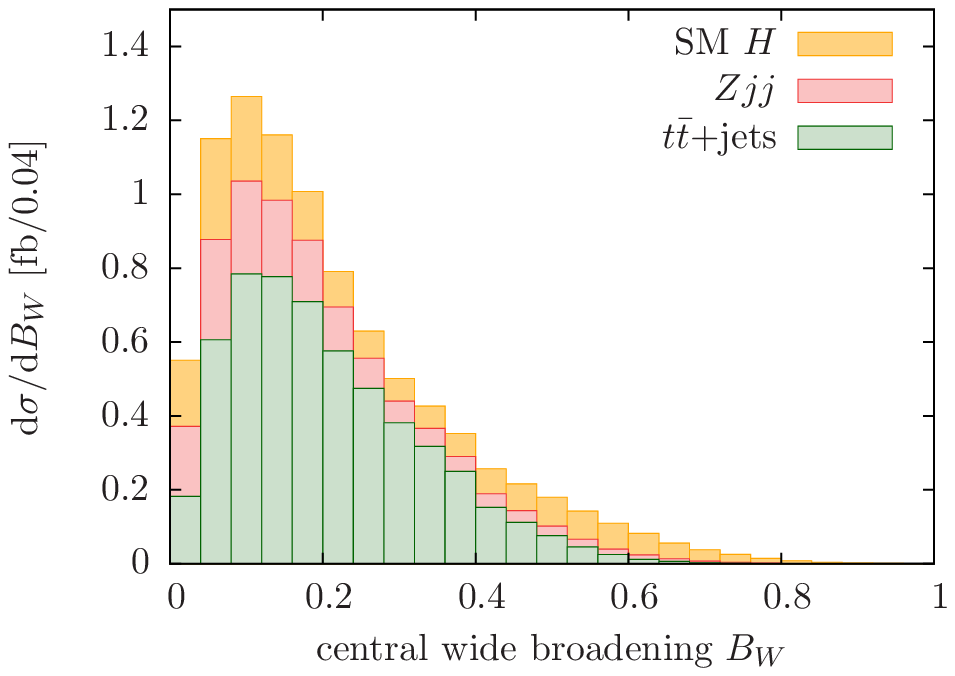}\hfill
 \includegraphics[width=0.32\textwidth]{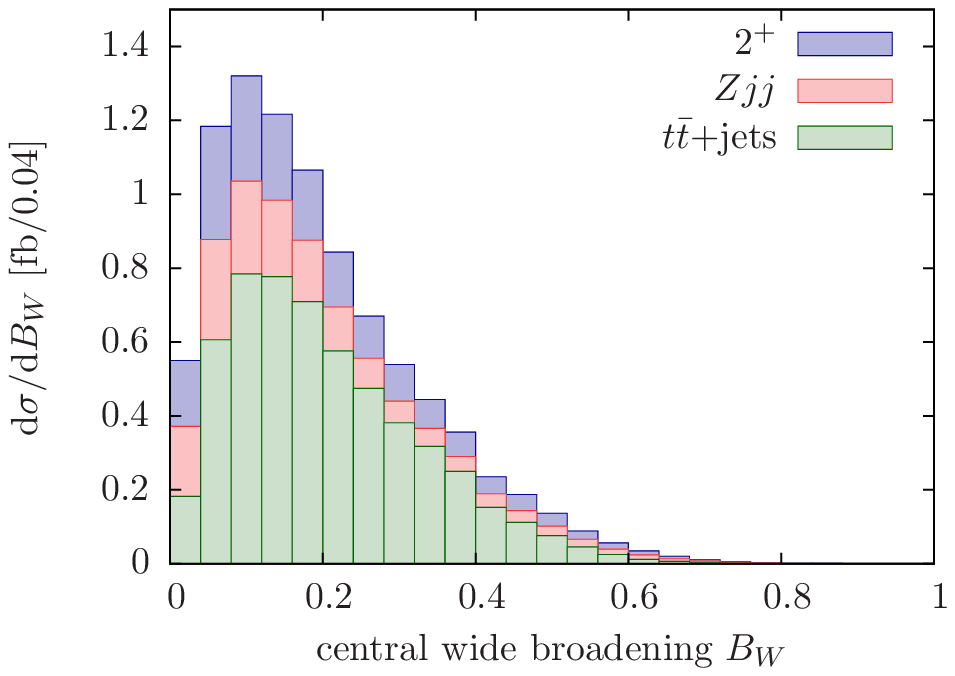}\hfill
 \includegraphics[width=0.32\textwidth]{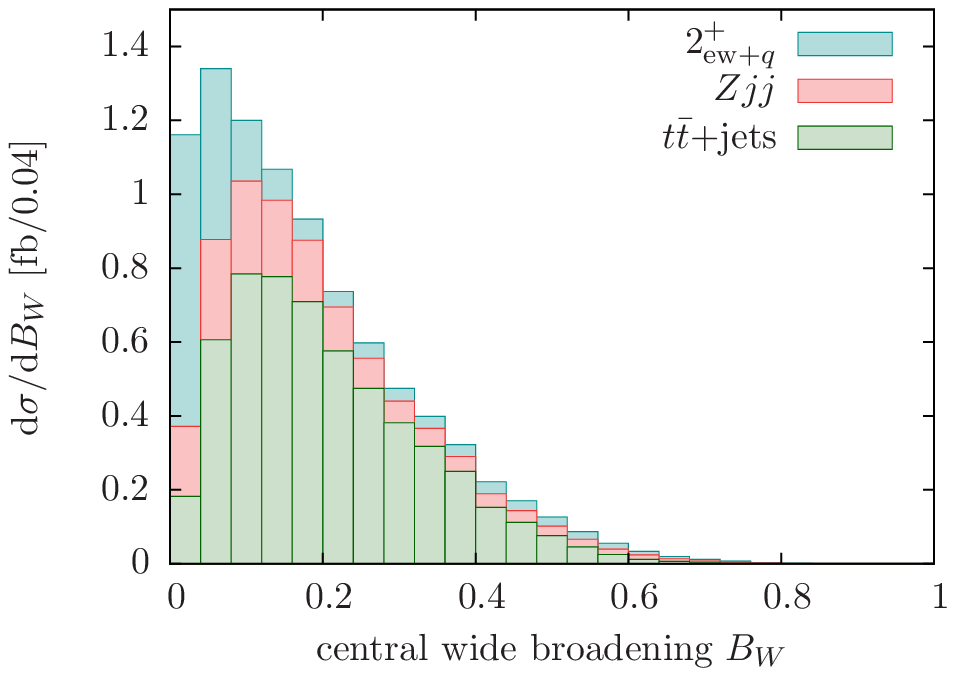}\\[0.6cm]
 \includegraphics[width=0.32\textwidth]{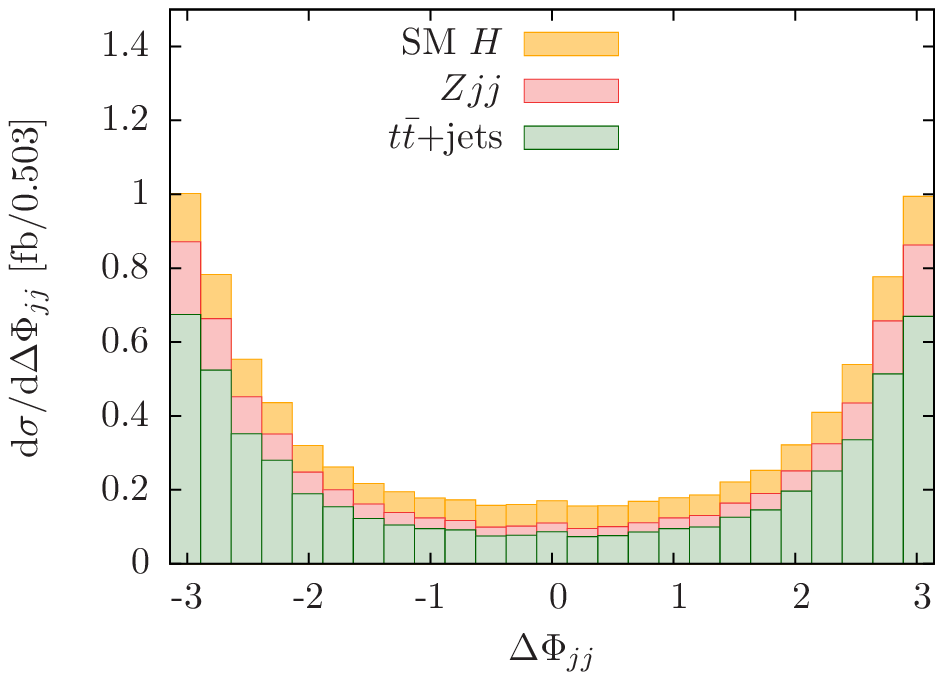}\hfill
  \includegraphics[width=0.32\textwidth]{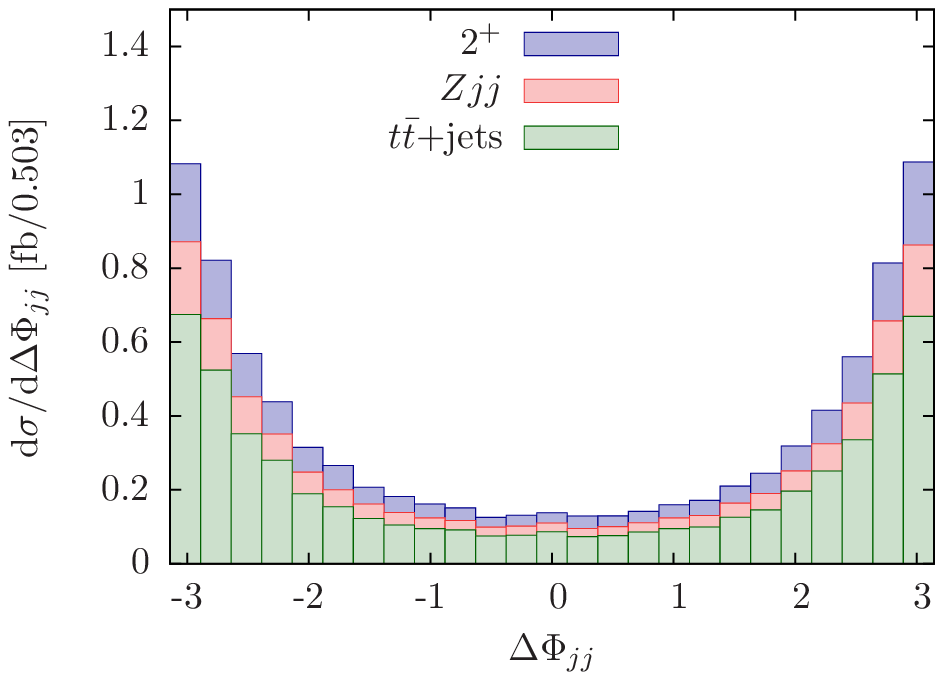}\hfill
  \includegraphics[width=0.32\textwidth]{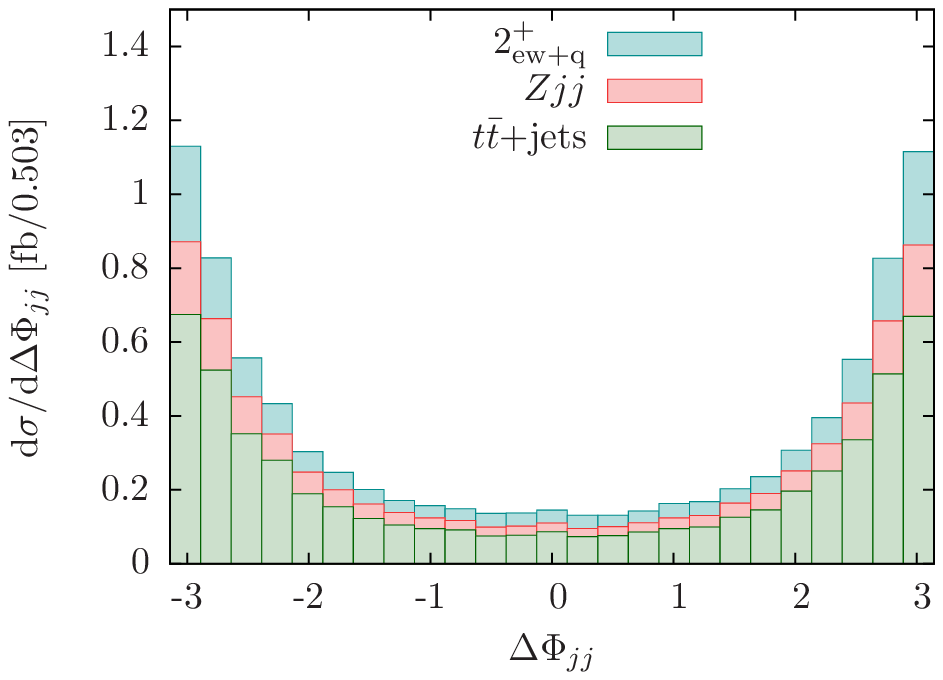}\\[0.2cm]
  \caption{\label{fig:evshapespilk} Event shape distribution
    calculated from jet constituents of selection (ii). We also show
    $\Delta\Phi_{jj}$.}
\end{figure*}
%

We apply a typical WBF selection~\cite{Plehn:2001nj,Englert:2012ct}
and cluster jets with the anti-kT algorithm~\cite{antikt} as
implemented in {\sc{FastJet}}~\cite{fastjet} with $D=0.4$ and define
jets with the thresholds
\begin{equation}
\label{eq:selectiona}
  p_{T,j}\geq 40~\gev,~{\text{and}}~|y_j|\leq 4.5\,.
\end{equation}
We impose an invariant mass cut on the two hardest tagging jets in the
event of
\begin{equation}
  \label{eq:selectiona1}
  m_{jj}=\sqrt{(p_{j,1}+p_{j,2})^2}\geq 600~\gev\,,
\end{equation}
and reconstruct the Higgs from taus with 
\begin{equation}
p_{\tau}\geq 20~\text{GeV and } |\eta_\tau|\leq 2.5
\end{equation} 
within a 50 GeV window around 125
GeV. The Higgs candidate has to fall between the tagging jets
\begin{equation}
    \label{eq:selection2}
  \min (y_1,y_2)<y_X < \max(y_1,y_2)\,.
\end{equation}
We further suppress the $t\bar t$+jets background by imposing a
central $b$ veto with an efficiency of $80\%$~\cite{atltag}. The
additional signal reduction due to mistagging is negligible within the
approximations we make. When normalizing {\it all signal samples to
  the SM Higgs cross section} after cuts ({\it{i.e.}} we treat the
$J=2$ hypotheses as Higgs-lookalikes) we have signal cross sections
$\sigma(X+2j)=3.82$~fb. The combined background is
$\sigma({\text{bkg}})=6.54$~fb.\footnote{The details of the cutflow
  are identical to Ref.~\cite{Englert:2012ct} and can be found in this
  earlier publication.}

\begin{figure*}[!t]
  \centering
  \subfigure[][~cut scenario (i)]{
    \includegraphics[width=0.43\textwidth]{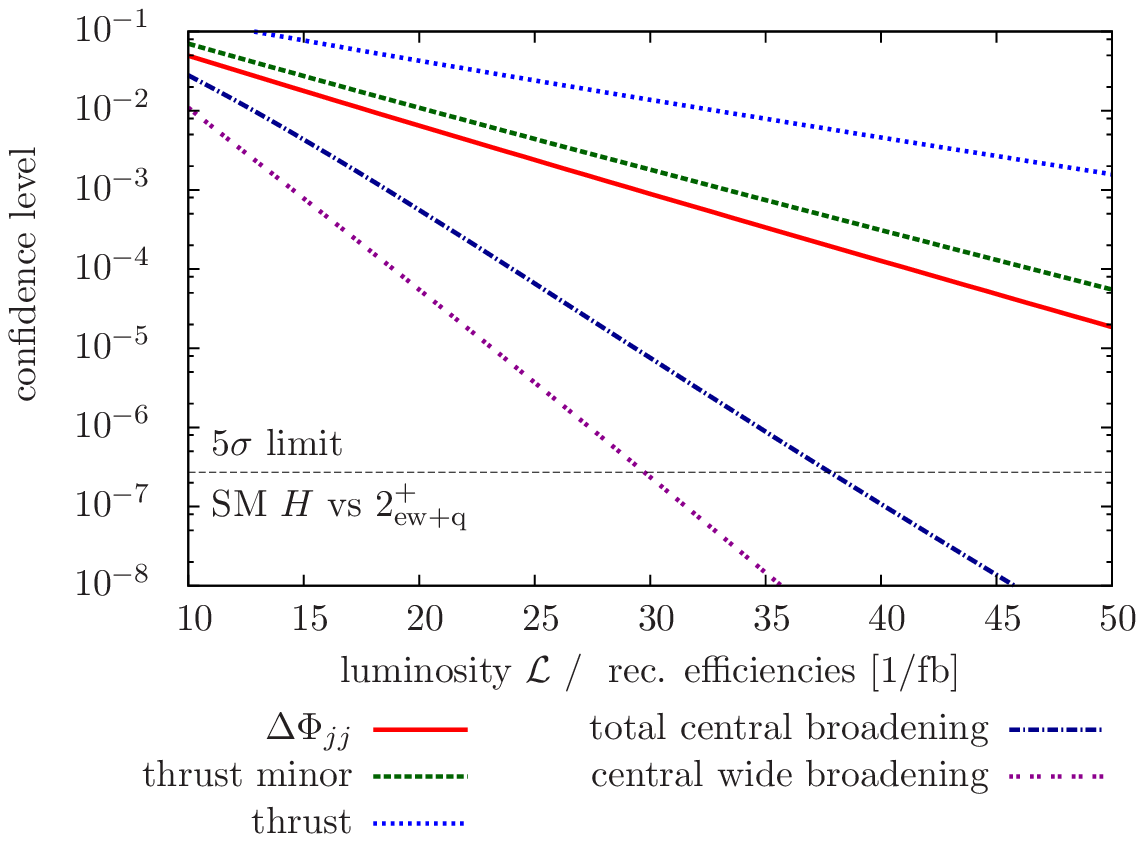}
  }\hspace{0.7cm}
  \subfigure[][~cut scenario (ii)]{
    \includegraphics[width=0.43\textwidth]{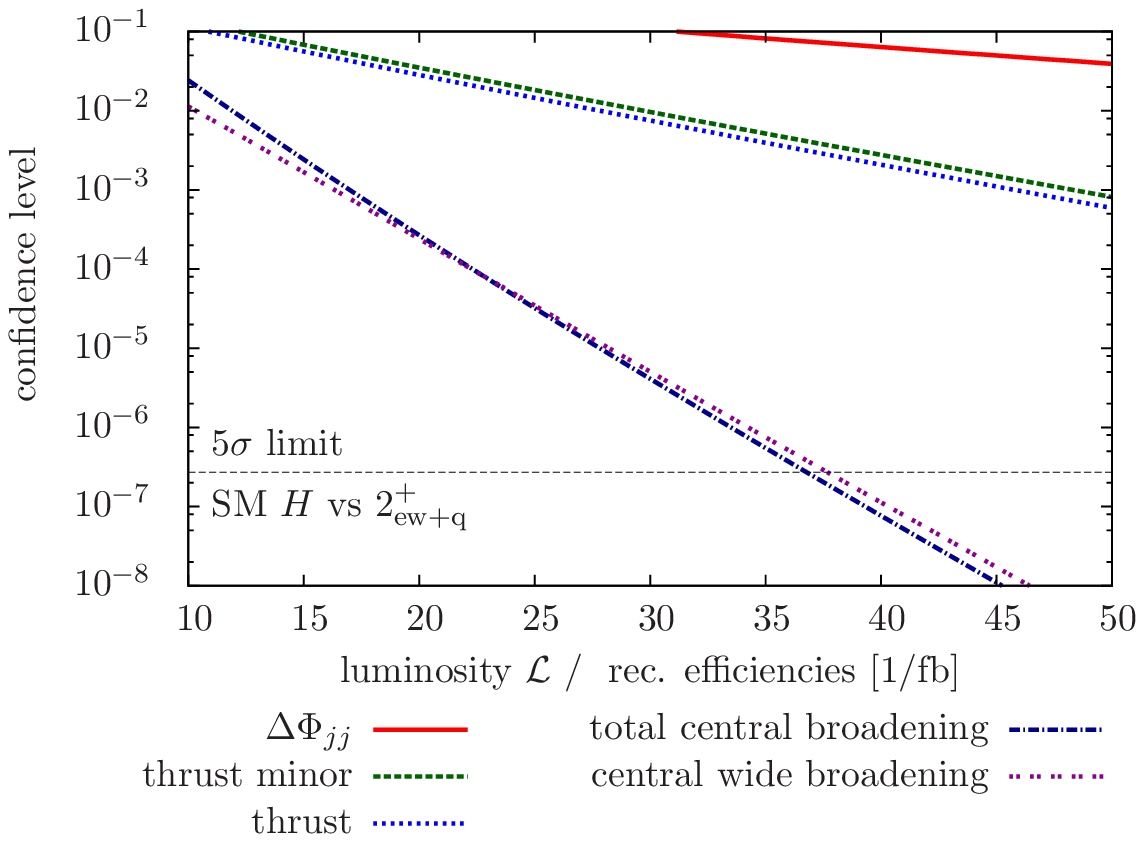}
  }
  \caption{\label{cls:inc} Result of the binned log-likelihood
    hypotheses test based on the input of selection~(i).}
\end{figure*}

\begin{figure*}[!t]
  \centering
  \subfigure[][~cut scenario (i)]{
    \includegraphics[width=0.43\textwidth]{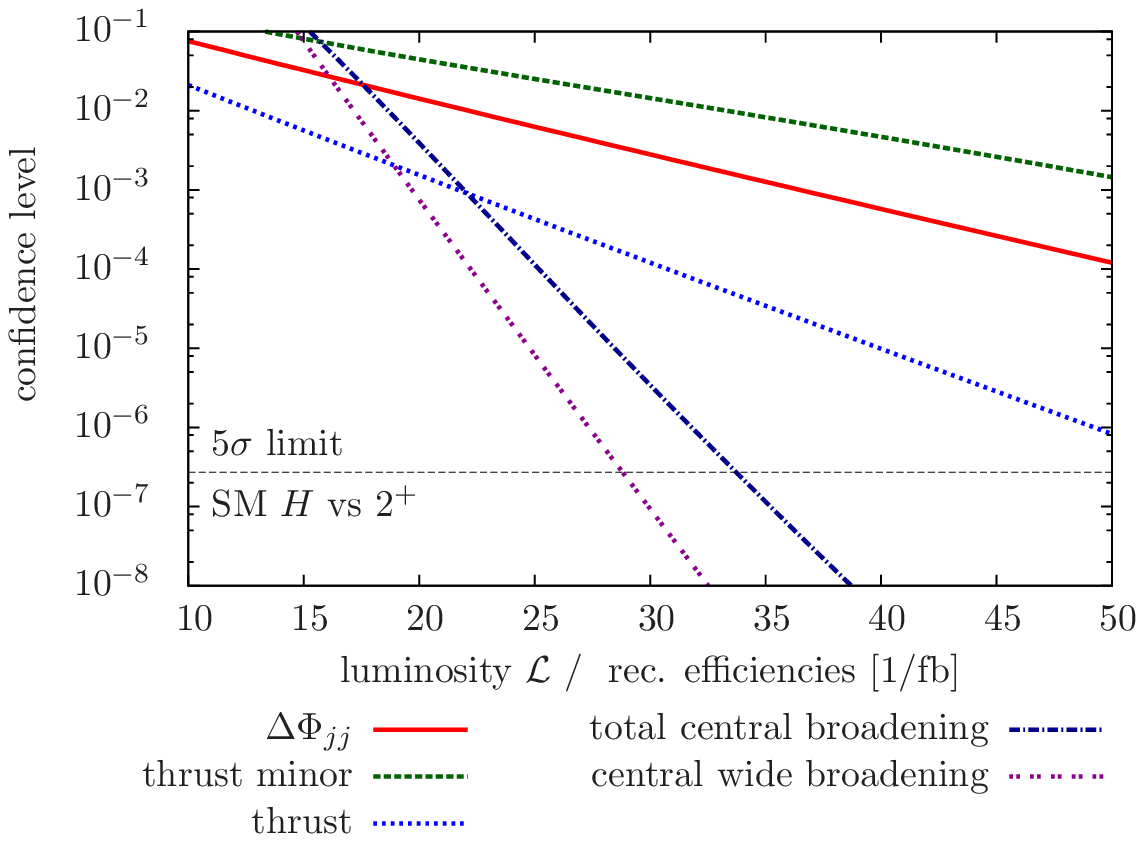}
  }\hspace{0.7cm}
  \subfigure[][~cut scenario (ii)]{
    \includegraphics[width=0.43\textwidth]{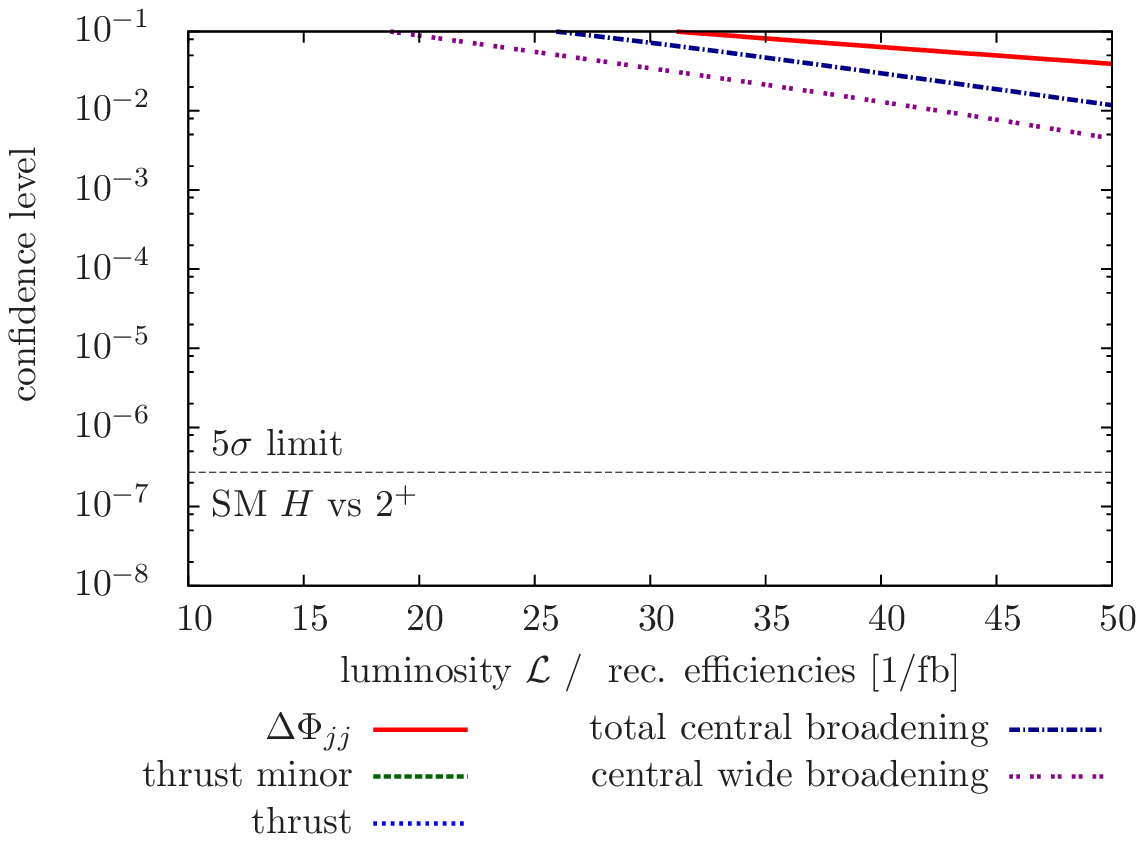}
  }
  \caption{\label{cls:pilk} Result of the binned log-likelihood
    hypotheses test based on the input of selection (ii).}
\end{figure*}

We proceed further by setting up two different track-selections that
eventually enter the evaluation of the considered event shape
observables. One of which is more robust against pile-up that can
cause issues when we want to study the global event properties
in the context of this paper.
\begin{enumerate}[(i)]
\item \label{en:i} For the events that pass the above selection
  criteria we feed all calorimeter hits with $p_T\geq 1$~GeV and
  $|\eta|\leq 4.5$ into the definition of the event shapes. This
  amounts to the most inclusive definition of the event shapes that is
  possible in the light of the above cuts. Selecting events according
  to the requirements Eq.~\gl{eq:selectiona}-\gl{eq:selection2} is
  already at odds with continuous globalness~\cite{evtshapes}, which
  guarantees good resummation properties \cite{evtshapes}. However,
  the used selection is the most inclusive possible in the light of
  unavoidable signal vs. background discrimination. To this end, we
  note that the analysis of Ref.~\cite{evtshapes} also shows that
  matched shower MCs reproduce the analytically resummed results well,
  so that we can expect our simulation to be under sufficient
  control. Quite obviously, this selection will be affected by pile-up
  activity.
\item The pile-up conditions for $\sqrt{s}=14$~TeV will need to be
  assessed when the LHC turns on again, but it can be expected that
  pile-up suppression in the central part of the detector is going to
  allow to lower jet thresholds in the rapidity region of the tracker
  $|\eta|\leq 2.5$~\cite{pusupp}. Currently, there is no tracking
  available for the more forward rapidity regions, so we will need to
  rely on hard jets to reduce in- and out-of time pile-up and
  underlying event.

  To reflect the effect of pile-up suppression to achieve a more
  robust definition of our observables we modify our event
  selection. We cluster jets as before, with the anti-kT algorithm and
  $D=0.4$, but this time we use the constituents of the jets obeying
  \begin{equation}
    p_{T,j}\geq 
    \begin{cases} 
      40~\text{GeV}\,, & 2.5< |\eta_j|\leq 4.5\\
      10~\text{GeV}\,, &  |\eta_j|\leq 2.5     
    \end{cases}
  \end{equation}
  as input for the event shapes instead of all particle tracks as
  considered in \gl{en:i}. This also allows one to enhance pile-up
  suppression by {\it{e.g.}} using the method of
  Ref.~\cite{pu2}. Furthermore, we explicitly require additional jet
  activity (specifically $n_j\geq 3$) which probes the spin structure
  induced radiation pattern. Since we are requiring at least three
  jets according to these modified criteria, the sensitivity we will
  find can be straightforwardly enhanced by including sensitivity from
  $\Delta\Phi_{jj}$, $\Delta\eta_{jj}$, $p_{T_j}$ (or equivalently
  $m_{jj}$) for the {\it exclusive} two jet category
  \cite{Englert:2012xt} in a hybrid observable approach.  We find
  cross sections for these cuts of $\sigma({\text{signal}})=1.89$~fb,
  while the background remains unchanged.\footnote{Note that this
    motivated central jet vetos~\cite{early,vetos} in the first
    place. The sensitivity we are going to find is lost in employing
    CJV-bases analysis strategies.}
\end{enumerate}

  For the spin 2 hypothesis 
  \begin{alignat}{5}
   &\mathcal{L}_2 =
    -g_1 \; G_{\mu\nu}T^{\mu\nu}_V
    -g_2 \; G_{\mu\nu}T^{\mu\nu}_G
    -g_3 \; G_{\mu\nu}T^{\mu\nu}_f \; , 
   \end{alignat}
   where $G_{\mu\nu}$ is the spin 2 resonance and $T^{\mu\nu}_{V,G,f}$
   is the energy-momentum tensor for the EW gauge bosons, gluons and
   fermions, we consider two representative
   scenarios~\cite{Englert:2012xt}.
\setlength{\leftmargini}{20pt}
\begin{description}
\item[$2^+:$] { The ordinary graviton-like tensor particle paradigm
    ({\it i.e.} $g_1=g_2=g_3= 1/\Lambda$), as considered in many other
    publications (see {\it e.g.} Ref.~\cite{more,Frank:2012wh}), has
    jet kinematics in the $X+2j$ final state that are close to the SM
    Higgs, once the additional selection cuts are
    imposed~\cite{Englert:2012xt}. The tagging jets are well-separated
    in $\eta$ and their $p_T$ distribution is not too different from
    the SM Higgs boson.}
\item[$2^+_{\rm{ew+q}}:$]{ We also study a model which has
    considerably harder jets while the WBF rapidity gap (and hence
    WBF-likeness) is still preserved. This specific model constrains
    the tensor couplings to weak bosons and fermions ({\it i.e.} $g_1=
    g_3 = 1/\Lambda$ and $g_2=0$).  This specific operator selection
    is therefore a less ``reasonable'' representative of a spin 2
    Higgs-lookalike.}
\end{description}
Our two choices will be clear from the discussion below, and are also
motivated by our findings for heavier Higgs-like particles in
Sec.~\ref{sec:future}.

The results of a number of event shape observables (for their
definition we refer the reader to the appendix and
Ref.~\cite{evtshapes}) are depicted in Fig.~\ref{fig:evshapes} for
selection~(i). This figure should be compared to
Fig.~\ref{fig:evshapespilk}, which displays the same distributions
subject to the modified requirements~(ii).

\begin{figure*}[!t]
  \centering
  \includegraphics[width=0.32\textwidth]{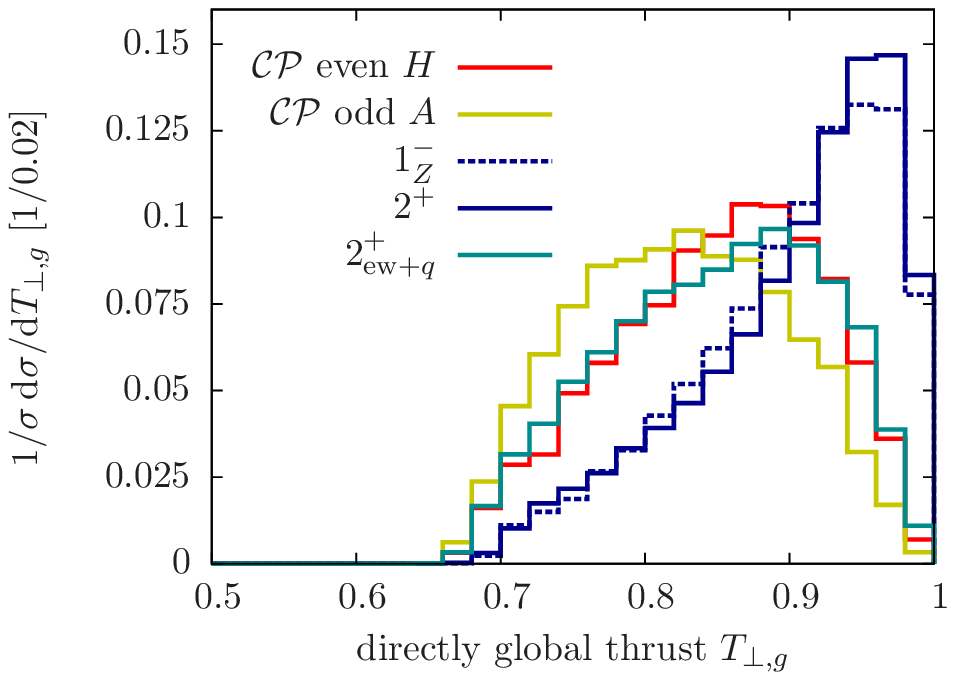}\hfill
  \includegraphics[width=0.32\textwidth]{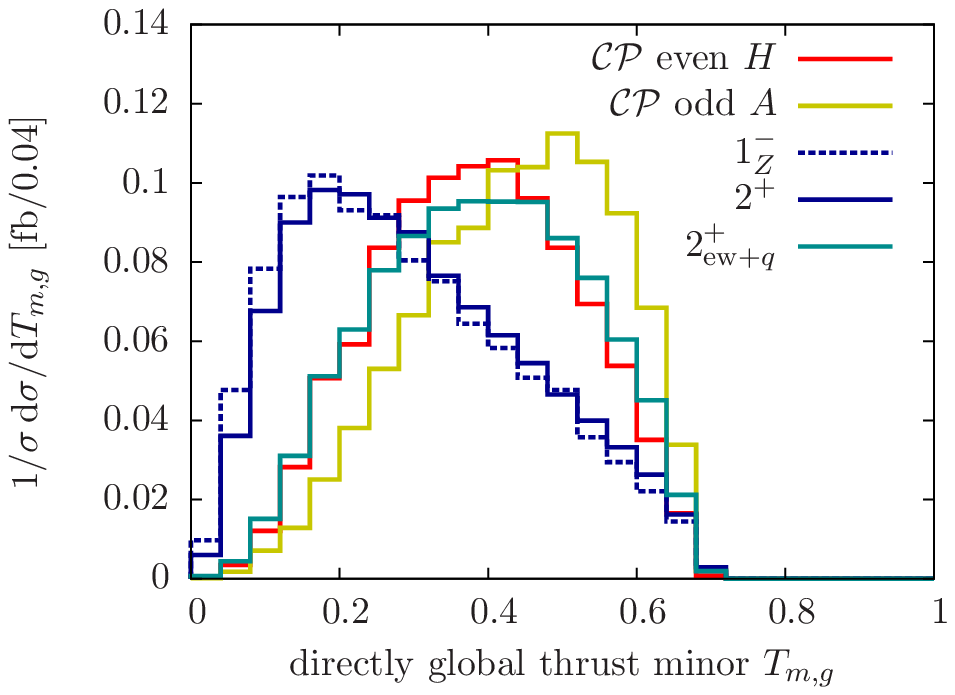}\hfill
  \includegraphics[width=0.32\textwidth]{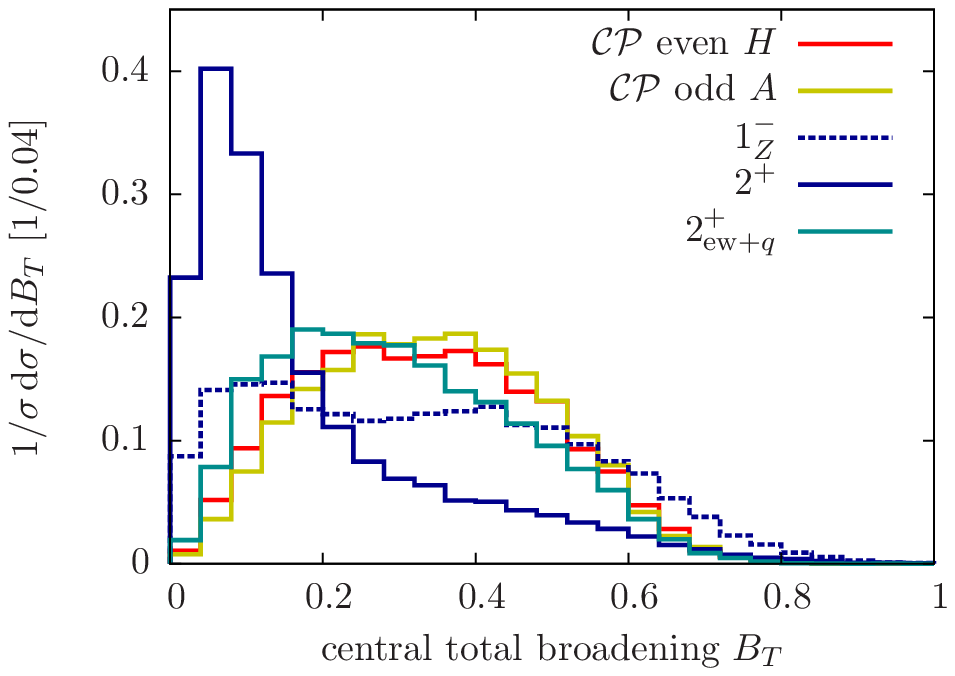}\\[0.4cm]
  \includegraphics[width=0.32\textwidth]{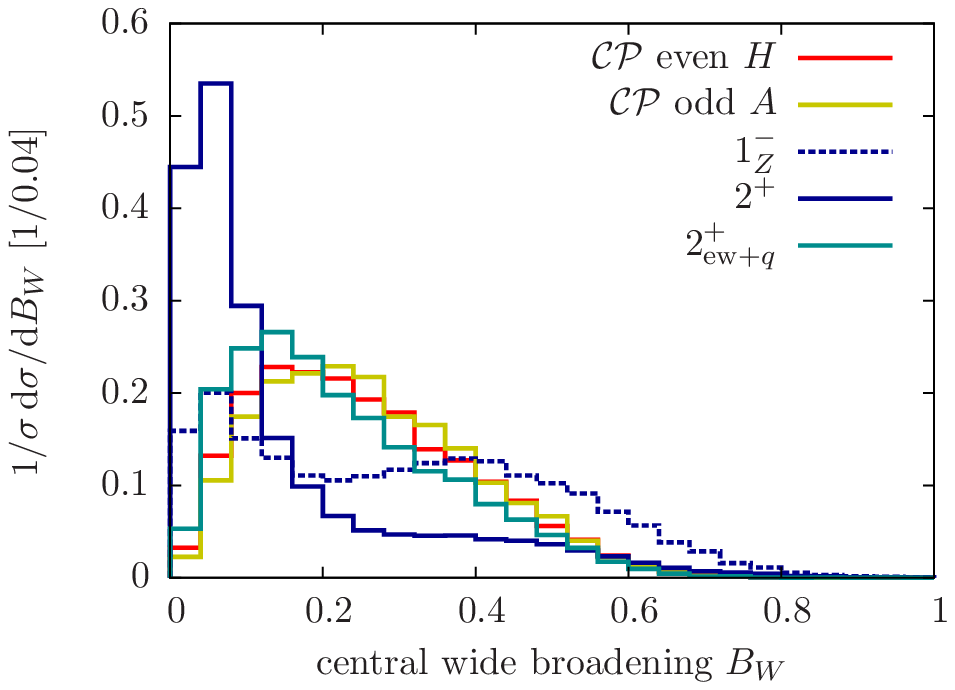}\hfill
  \includegraphics[width=0.32\textwidth]{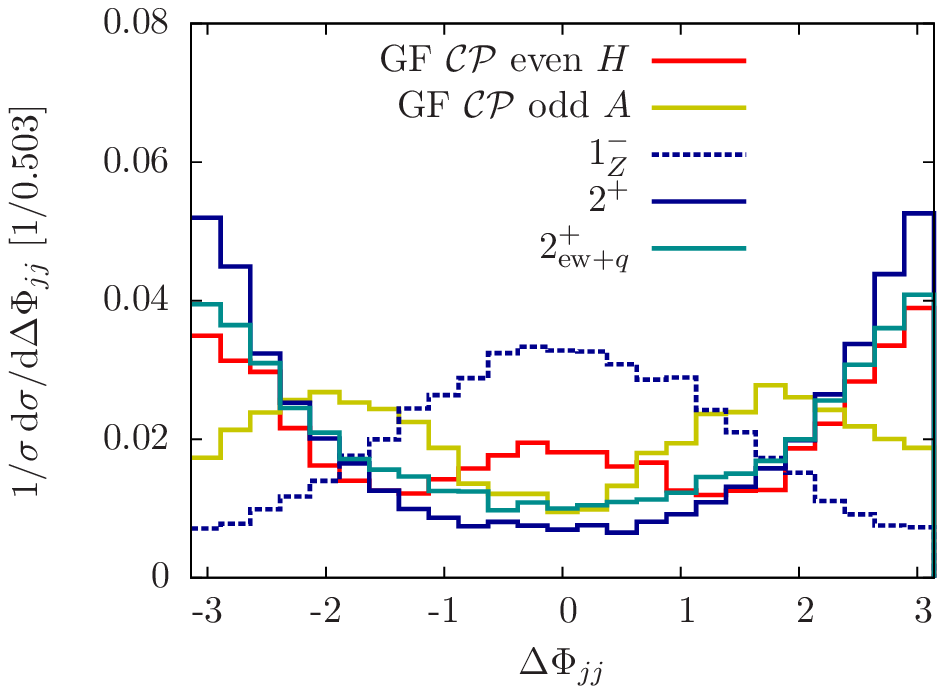}\hfill
  \parbox{0.32\textwidth}{
    \vspace{-3cm}
    \hspace{0.8cm}
    \parbox{0.28\textwidth}{ \caption{\label{fig:heavy} Event shape and $\Phi_{jj}$
        distributions for selection (i) and $m_X=300$~GeV.}
    }
  }
\end{figure*}

To quantify the statistical discriminative power of the event shape
observables we perform a binned log-likelihood hypothesis
test~\cite{llhr} in Figs.~\ref{cls:inc} and \ref{cls:pilk}; this
provides a statistically well-defined estimate of the luminosity (upon
dividing out all reconstruction efficiencies) that is required to
reject the spin 2 hypotheses at the $5$~sigma level using the CL$_S$
method~\cite{Read:2002hq}.

The results of this analysis are shown in Fig.~\ref{cls:inc} for the
$2^+_{\text{ew+q}}$ and in Fig.~\ref{cls:pilk} for the $2^+$ cases. As
already expected from Figs.~\ref{fig:evshapes} and
\ref{fig:evshapespilk}, the broadening observables perform
best. Depending on the specific scenario, these observables are robust
against pile-up as discussed in (ii). Fig.~\ref{cls:pilk}, however,
also shows that, when the jet kinematics become SM-like, this will be
reflected in a lower sensitivity of the event shapes to the involved
spin hypothesis. This especially holds when discriminative power at
smaller broadening is lost due to soft radiation not taken into
account for selection~(ii)~vs.~(i). This also explains our initial
choice of the spin 2 hypotheses: $2^+$ is characterized by soft
radiation and therefore suitable to be studied using event shape
observables. We find broadening observables to provide the strongest
statistical sensitivity. However, while this model can be formidably
constrained using event shapes if pile-up is under sufficient control,
{\it i.e.}, when the actual selection can be chosen closer to~(i), the
discriminative power of the broadening observables is severely reduced
for selection~(ii). On the other hand, $2^+_{\text{ew+q}}$ which has a
slightly harder spectrum is robust in our comparison~(i)~vs~(ii) and
the event shape observables provide a statistically appealing
single-valued discriminant.

\section{Spin discrimination of future Higgs-like resonances}
\label{sec:future}
Let us finally comment on the prospect of using the methods of the
previous section also in the context of spin analyses of Higgs-like
states that might be discovered in the future with a heavier
mass. This is not immediately clear since the higher mass scale
implies a different (soft) radiation pattern. As a representative
example we discuss $m_X\simeq 300$~GeV.

In general we can expect relatively small couplings of this additional
state to the electroweak gauge bosons $Z$ and $W$, as current
measurements seem to suggest that unitarity cancellations, which
characteristically determine the couplings of additional massive
scalars with corresponding couplings, are saturated by the 125~GeV
state. The standard technique in $X\to ZZ$ \cite{cp,chack} might hence
not be applicable and an investigation of the $X+2j$ final state could
well be the only phenomenologically available channel to constrain the
spin and {\cal{CP}} structure of such a discovery.

We consider these reasons as enough motivation to limit ourselves for
scalar boson candidates to the gluon fusion channel
Fig.~\ref{fig:feyngraphs} (a). For the spin 2 candidates we will again
adopt the scenarios of the previous section, which will have quite
different phenomenology as compared to the $m_X=125$~GeV case.

For spin 1 candidates our above arguments constrain the interactions
of copies of the SM gauge bosons. The phenomenology of a Kaluza-Klein
excitation spectrum as encountered in {\it e.g.} warped extra
dimensions (and their dual interpretation as vectorial and axial
vector resonances of a strongly-interacting sector) is therefore
heavily suppressed in the SM vector boson final states. There is an
exception to the unitarity argument which are $Z'ZZ$ interactions as
determined in the generalized Landau Yang theorem \cite{gly}. The
structure of the interaction vertices does not introduce an
energy-dependent unitarity violation and hence, is not constrained by
current measurements. We include this interaction to model a WBF
(Fig.~\ref{fig:feyngraphs} (b)) spin 1 candidate $J(X)=1^-_Z$.

\begin{figure*}[!t]
  \centering
  \includegraphics[width=0.32\textwidth]{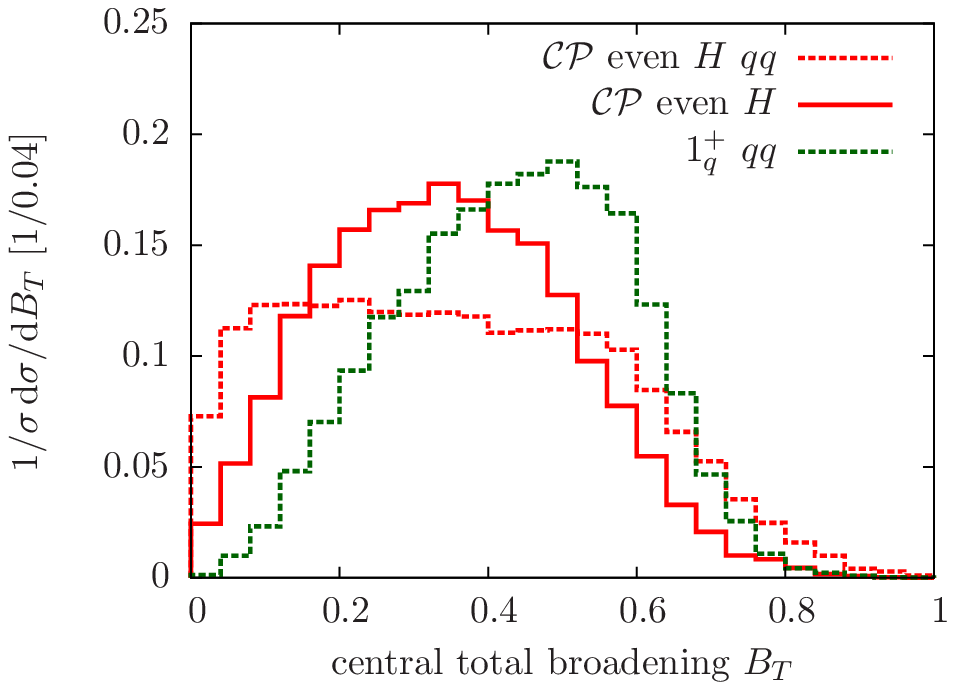}\hfill
  \includegraphics[width=0.32\textwidth]{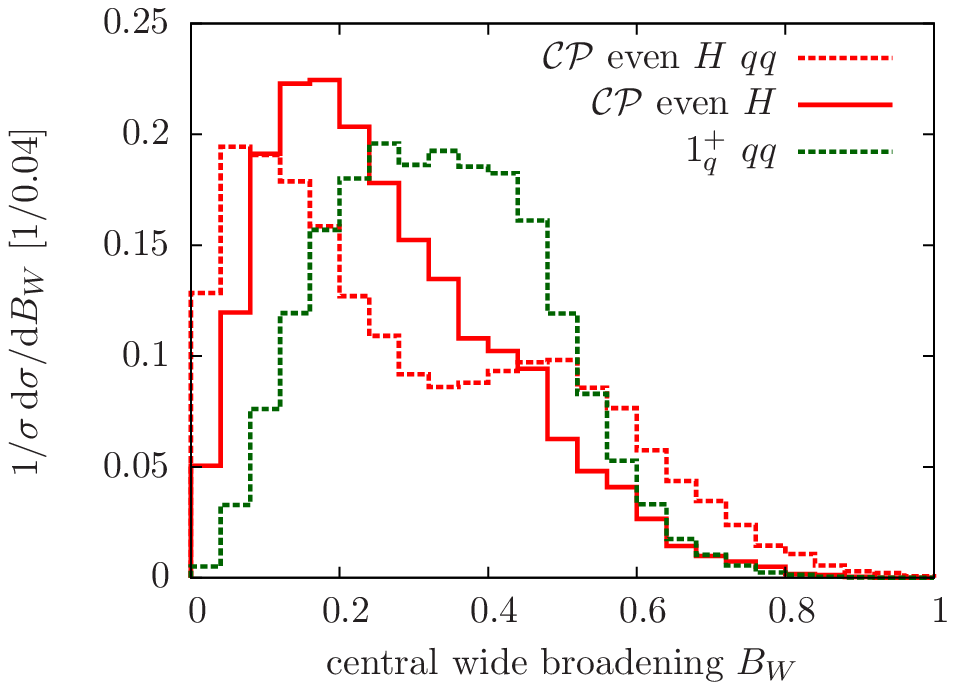}\hfill
  \includegraphics[width=0.32\textwidth]{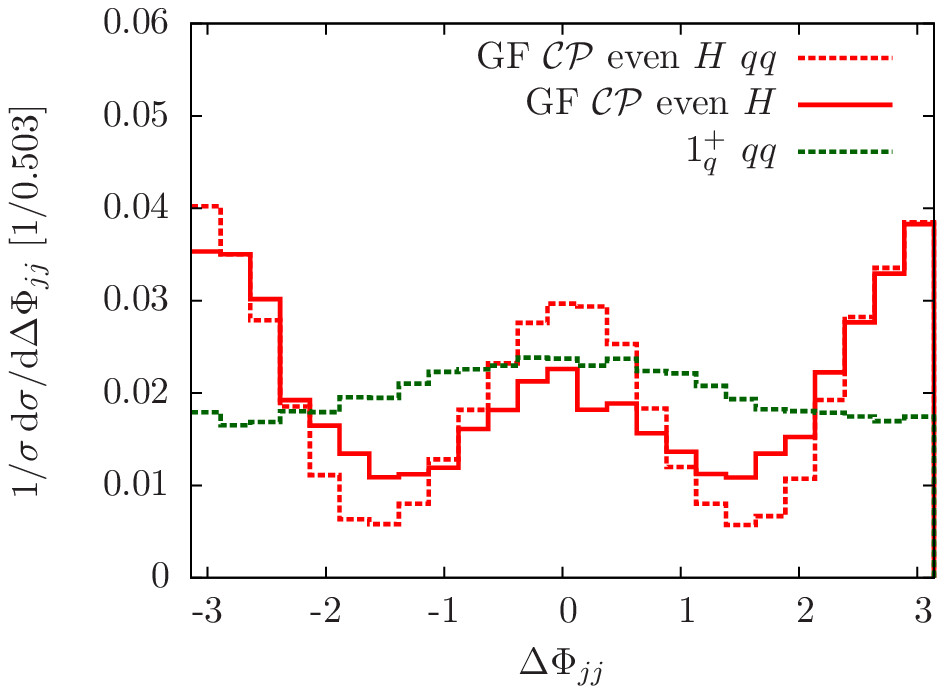}\\[0.2cm]
  \caption{\label{fig:heavyb} Event shape comparison for the SM Higgs and $1^+_q$
  for the $qq$-induced channels including the full mass dependence $m_X=300$~GeV.}
\end{figure*}

Gluon-fusion contributions for spin 1 degrees of freedom analogous to
Fig.~\ref{fig:feyngraphs} (a) are more difficult to model. Furry's
theorem \cite{furry} guarantees the exact cancellation of vector
current from $J^{\cal{CP}}(X)=1^-$ hypothesis in $gg\to X$.  Axial
vector currents still have to obey the Landau Yang theorem
\cite{landauyang}.  This renders an observation of prompt gluon fusion
impossible; on-shell production exactly vanishes and gluon fusion
becomes a function of the $J^{\cal{CP}}(X)=1^+$ particle's width and
the virtuality of the gluon. These small effects are at odds with
conventional bump searches and leave gluon fusion, as depicted in
Fig.~\ref{fig:feyngraphs} (a), as the only production mechanism when
such a state has suppressed couplings to the SM $Z$'s (these couplings
are again determined by the generalized Landau Yang theorem). While
the particle $X$, in Fig.~\ref{fig:feyngraphs} (a), can be considered
on-shell for resonance-driven searches, the $t$ channel gluons are
always off-shell: this enables $J^{\cal{CP}}(X)=1^+$ production via
gluon fusion (see also Ref.~\cite{vanderBij:1988ac}). For the moment
we are not interested in a survey of the effects of $d>6$ operators
that are involved in these interactions \cite{Buchmuller} on the
events' energy momentum flow. We however note that different effective
operators will contribute to the gluon-gluon, gluon-quark, and
quark-quark channels.

Instead, we will model axial vector particles in gluon fusion plus two
jets by introducing a doublet of heavy fermions, which couple to the
axial vector boson with couplings chosen such that anomaly
cancellation is manifest. We keep the full mass dependence by
simulating $q q'\to 1^+_{\text{q}} + q q'$ events with a modified
version of {\sc{MadGraph}} v4.4 \cite{mad4} that links a customized
one-loop capable {\sc{Helas}} \cite{helas} library. To gain a
qualitative picture we compare the energy momentum flow of this model
against the corresponding full one-loop SM Higgs events $qq'\to
0^+_{\text{SM}} + qq'$. The gluon-induced channels will populate more
the central region, but do not change the overall
picture.\footnote{The quark-gluon and gluon-gluon- induced channels do
  not introduce a different $\Delta\Phi_{jj}$ radiation pattern for
  instance \cite{Klamke:2007cu,paco}, Fig.~\ref{fig:heavyb}.}

The (normalized) results are presented in Fig.~\ref{fig:heavy} for the
identical jet cut setup of Sec.~\ref{sec:analysis}. We do not include
the backgrounds as these depend on the specific decay channel in which
such a future resonance will be discovered. Typical QCD background
suppression will however always be centered around the cuts of the
previous section, independent of the specific exclusive decay channel
of $X$. From the shown distributions it is clear that there is
substantial discriminative power in separating the scalar options from
$1^-_Z$ and $2^+$ in the event shape observables. A combination with
ordinary jet-based observables such as $\Delta\Phi_{jj}$ will serve to
discriminate these options further for tighter selections if feasible.

In Fig.~\ref{fig:heavyb} we finally show the comparison of the quark
channels for the $1^+_\text{q}$ vs. $1^+_{\text{SM}}$, which also
provides insights how different partonic channels (and hence effective
operators) will influence our findings. Indeed the shapes are rather
identical to Fig.~\ref{fig:heavy} for the scalar boson; we can
therefore expect that the event shapes also serve to discriminate
between $0^+$ and $1^+$ for various spin template combinations, beyond
the approximations we have made.  Note also that our spin 2 hypotheses
behave completely opposite compared to the $m_h=125$~GeV case due to
the changed momentum dependence of the cross section on the tagging
jets. In this sense, $2^+_{\text{ew+q}}$ provides a better alternative
hypothesis than $2^+$ when such a measurement is performed in the
future.

\section{Summary and Conclusions}
The recent discovery of a Higgs-like particle at the LHC and further
measurements of it seem to suggest that we have indeed discovered a
particle which is consistent with the $J^{\cal{CP}}(X)=0^+$ SM Higgs
boson prediction. Analyses with increased statistics across many
different channels will allow to answer the $J^{\cal{CP}}$ question
more reliably. The $pp\to X+2j$ mode, when analyzed in inclusive
selections, provides a valuable channel to discriminate between
different spin (and \cp) hypotheses when the events' global QCD
energy-momentum flow pattern is analyzed. The latter is most
efficiently captured in event shape distributions. While thrust
provides a straightforward handle to discriminate discrete
\cp~values~\cite{Englert:2012ct}, the broadening observables reflect
the spin-induced radiation patterns. Issues that may arise from
challenging pile-up conditions can be counteracted with adopted
definitions of the event shape observables and hybrid
exclusive/inclusive definitions of the employed single valued
discriminants. Depending on the spin 2 scenario (no spin 2 scenario is
theoretically motivated but merely invoked as an alternative
hypothesis to be excluded) we find large discriminative power in the
accompanied energy momentum flow. This generalizes the results of
Refs.\cite{Frank:2012wh,Englert:2012xt,Djouadi:2013yb,Englert:2012ct}.
Pile-up, as for many analyses, can become a challenge of the discussed
analysis strategy to the point where discriminative power in all
collider observables is lost in the $X+2j$ final state. This again
highly depends on the chosen hypothesis.

Given the consistency of the observed cross sections in $pp \to X\to
ZZ,W^+W^-$ with the SM Higgs boson, it is likely that spin analyses of
an additional resonance as predicted by many BSM scenarios cannot be
straightforwardly performed in the $X\to \gamma\gamma,ZZ$ channels. In
this case an event shape based analysis of the QCD energy momentum
flow might be crucial since it does not rely on a particular exclusive
final state decay of $X$. Indeed, we find significant discriminative
power of the event shape observables for heavier ``Higgs'' masses,
which allows to discriminate various $J^{\cal{CP}}$ hypotheses in
combination with exclusive $2$-jet measurements in the same channel
\cite{Englert:2012xt}. As shown in this work, the advantages of event
shape-based analyses are not limited to the study of pure QCD events
but clearly generalize to the interplay of QCD with the (BSM)
electroweak sector.

\bigskip

\noindent {\bf{Acknowledgments}} --- We thank Andrea Banfi, Gavin
Salam, and Giulia Zanderighi for providing the {\sc{Caesar}} event
shape library.
C.E. acknowledges funding by the Durham International Junior Research
Fellowship scheme.


\appendix

\section{Definitions of the studied event shapes}

Event shapes are widely used observables to investigate geometrical
properties of particle collisions at lepton and hadron
colliders~\cite{Brandt:1964sa,thrustcms,broadening,thrustexi,thrust},
which can be described to very high theoretical accuracy, see {\it
  e.g.} \cite{evtshapes,thrust}. At hadron colliders one typically
defines the observables in the beam transverse plane.
{\it{Transverse thrust}} is therefore defined as the maximization
procedure in the transverse plane
\begin{equation}
  \label{eq:thrust}
  T_{\perp,g} = \max_{{\vec{n}}_T} \frac{ \sum_i |{\vec{p}}_{\perp,i} \cdot
    {\vec{n}}_T|}{\sum_i | {\vec{p}}_{\perp,i}|},\quad |{\vec{n}}_T|=1\,,
\end{equation}
where $p_{T,i}$ denotes the transverse momentum of the track~$i$.  The
transverse thrust value of circularly symmetric event is
$T_{\perp,g}=2/\pi\simeq 0.64 $, while an ideal alignment is
characterized by $T_{\perp,g}= 1$.

As a result of the maximization procedure we obtain the transverse
thrust axis ${\vec{n}}_T$ which enters the definition of transverse
{\it{thrust~minor}}
\begin{equation}
  \label{eq:thrustmin}
  T_{m,g} = \frac{ \sum_i |{\vec{p}}_{\perp,i} \times
    {\vec{n}}_T|}{\sum_i | {\vec{p}}_{\perp,i}|} \,,
\end{equation}
which measures the energy-momentum flow perpendicular to the
transverse thrust axis.

Observables that are particularly helpful in the context of spin
analyses are broadening observables~\cite{broadening}. For their
definitions we first specify a central region, $C$, in terms of
pseudorapidity; here $C$ corresponds to $|\eta| \leq 4.5$. Then we
split this region according to transverse thrust axis
\begin{subequations}
  \label{eq:wrapbroad}
  \begin{equation}
    \text{region~} \begin{matrix} {C_U}\\ {C_D} \end{matrix} \quad  {\vec{p}}_{\perp,i}\cdot {\vec{n}}_{T}
    \gtrless 0
  \end{equation}
  and subsequently compute the weighted pseudorapidity and azimuthal
  angle
  \begin{multline}
    \eta_{\sigma}=\frac{ \sum_{i} |{\vec{p}}_{\perp,i}|
      \, \eta_i }{
      \sum_{i} |{\vec{p}}_{\perp,i}|} \,,
    \quad
    \phi_{\sigma}=\frac{ \sum_{i} |{\vec{p}}_{\perp,i}|
      \, \phi_i }{
      \sum_{i} |{\vec{p}}_{\perp,i}|}\,,\quad\sigma=C_U,C_D.
  \end{multline}
  The broadening of the above regions is then defined as
  \begin{multline}
    B_{\sigma}={1\over 2Q_{T} }\sum_{i \in \sigma}
    |{\vec{p}}_{\perp,i}|\sqrt{(\eta_i -
      \eta_{\sigma})^2+(\phi_i-\phi_{\sigma})^2}\,,\\\sigma=C_U,C_D
  \end{multline}
  with $Q_{T} = {\sum_i | {\vec{p}}_{\perp,i}|}$.  The {\it{central
      total broadening}} and {\it{central wide broadening}} observables are
  \begin{equation}
    \label{eq:broadening}
    \begin{split}
      \text{central total broadening: }&
      B_{T}=B_{C_U}+B_{C_D}\,, \\
      \text{central wide  broadening: }&
      B_{W}=\max\left\{B_{C_U},B_{C_D}\right\}\,.
    \end{split}
  \end{equation}
\end{subequations}


\end{document}